\documentclass[useAMS,usenatbib]{mnras}

\usepackage{newtxtext,newtxmath}

\usepackage[T1]{fontenc}
\usepackage[normalem]{ulem}

\DeclareRobustCommand{\VAN}[3]{#2}
\let\VANthebibliography\thebibliography
\def\thebibliography{\DeclareRobustCommand{\VAN}[3]{##3}\VANthebibliography}

\usepackage{graphicx}	
\usepackage{amsmath}	

\usepackage[version=3]{mhchem}

\title[C/O in the inner discs]{Metamorphoses of carbon and oxygen in protoplanetary discs: how chemistry and radial drift transform inner disc C/O ratios}

\author[Molyarova et al.]{
Tamara Molyarova$^{1}$\thanks{E-mail: t.molyarova@leeds.ac.uk},
Richard A. Booth$^{1}$,
Catherine Walsh$^{1}$
\\
$^{1}$School of Physics and Astronomy, University of Leeds, Leeds LS2 9JT, UK
}

\date{Accepted XXX. Received YYY; in original form ZZZ}

\pubyear{\the\year{}}

\begin{document}
\label{firstpage}
\pagerange{\pageref{firstpage}--\pageref{lastpage}}
\maketitle

\begin{abstract}
The chemical composition of a protoplanetary disc is sensitive to its thermal structure and dust properties, and can provide insights into the disc evolution.
Recent observations with the James Webb Space Telescope (JWST) reveal correlations of the inner disc compositions with disc size, accretion rate and stellar mass, explained by the key role of dust radial drift in redistributing primordial volatiles.
We explore how chemical reactions change the composition of ices carried with pebbles and how they affect the inner disc C/O ratios in a disc around a solar mass star.
We consider different drift efficiencies set by dust fragmentation velocity and include dust traps at different locations. We vary the incident cosmic ray ionisation rate $\zeta$ and the efficiency of cosmic ray dissociation of ices, and consider the effect of carbon grain destruction. We find that methane depletion within $<1$\,Myr prevents the delivery of carbon-rich gas to the inner disc and yields $\mathrm{C/O} \lesssim1$ for $\zeta\geq10^{-17}$\,s$^{-1}$. 
Dust traps collect water and carbon-rich ices formed via methane destruction, further lowering the inner disc metallicity and C/O ratio. Cosmic-ray driven photodissociation of ices can convert water to \ce{O2} and carbon-bearing molecules to \ce{CO}, allowing ices to escape the trap if $\gtrsim10 \%$ of the dissociated products can participate in surface reactions.
We discuss the observational implications and conclude that cosmic rays and their effect on ices are the key factors that determine the impact of chemistry on the inner disc composition.

\end{abstract}




\section{Introduction}

The chemical composition of protoplanetary discs sets the environment for planet formation. The composition is determined by the combination of dust growth and chemical processes in the gas and on dust grains, and bears the imprint of global evolution of the disc physical structure. Thus, observations of the molecular and elemental composition of protoplanetary discs can probe the physical properties of protoplanetary discs and offer an insight into their evolution.

Both chemical and physical processes affect the disc composition; the former can repartition key elements between molecular carriers, while the latter redistributes them between the gas and ice phases, as well as between different disc regions. A variety of protoplanetary disc models consider chemical reactions and the evolution of dust and gas properties either separately or in combination, to simulate the elemental and molecular composition of planet-forming material. The assumption of a quasi-static disc structure allows one to thoroughly model chemical evolution, accounting for thousands of reactions in vertically and radially resolved simulations \citep[e.g.][and many others, see \citet{2013ChRv..113.9016H} for a review]{2002A&A...386..622A,2009ApJ...693.1360W,2009A&A...501..383W,2013ApJ...766....8A,2014Sci...345.1590C,2015A&A...582A..88W,2016A&A...595A..83E,2017A&A...607A..41K,2019ApJ...885..146R,2019ApJ...877..131S}. Some of these models include turbulent mixing \citep{2011ApJS..196...25S,2013ApJ...779...11F} or temporal evolution of physical structure \citep[e.g.][]{2014MNRAS.445..913D,2016MNRAS.462..977D,2017A&A...604A..15R,2018A&A...613A..14E,2019ApJ...883..197W,2021MNRAS.500.4658M}. 

Another class of models focuses on the dust evolution and transport and only includes freeze-out chemistry \citep{2016ApJ...831L..19O,2016ApJ...833..203P,2017MNRAS.469.3994B,2021A&A...654A..71S,2023ApJ...954...66K,2024A&A...686L..17M,2025PASA...42....4M,2025A&A...694A..79S,2025MNRAS.tmp.1736W}. These models typically consider the disc midplane only and allow to capture the redistribution of elements, particularly in the radial direction. Some models are focused on the vertical dust transport, with chemistry \citep{2022ApJ...927..206V} or without \citep{2016ApJ...833..285K,2018ApJ...864...78K}. 
Finally, some global models combine aspects of both chemistry and dust evolution \citep[e.g.,][note that the listed works have varying degrees of chemical complexity]{2007A&A...463..369T,2017MNRAS.472..189I,2017MNRAS.469.3910C,2019MNRAS.487.3998B,2020ApJ...899..134K,2025A&A...701A.194P,2025A&A...701A.239S}. Another type of model are those that post-process the physical evolution of the disc and follow the chemical evolution of a gas element \citep{2017ApJ...835..162X,2020ApJ...890..154P,2022A&A...667A.160E}.

The relatively colder outer regions of protoplanetary discs ($\gtrsim10$\,au) are observed at millimetre wavelengths in rotational transitions of gas-phase molecules by Atacama Large Millimeter/submillimeter Array (ALMA) or Northern Extended Millimeter Array (NOEMA). Observations indicate that the gas in the outer disc is often carbon-rich, with super-solar C/O ratios, often $\gtrsim1$~\citep{2016A&A...592A..83K,2016ApJ...831..101B,2018ApJ...865..155C,2021ApJS..257....7B,2023A&A...670A..12S,2019A&A...631A..69M}. The modelling of the observed molecular abundances indicates that the elevated C/O ratio is likely a result of the vertical transport of ice-covered grains between irradiated and non-irradiated layers leading to chemical processing of CO by UV radiation and subsequent formation of hydrocarbons \citep{2016ApJ...831..101B,2022ApJ...927..206V}. Another factor is the depletion of carbon and oxygen through radial dust transport inwards, which can be affected by the presence of ring substructures and their position relative to the CO snowline \citep{2021A&A...653L...9V,2022A&A...660A.126S}.

The composition of hotter inner discs can be probed through near- to mid-infrared observations, a capability enhanced by JWST/MIRI \citep[e.g.,][]{2023ApJ...957L..22B,2023ApJ...947L...6G,2024A&A...689A..85G,2024ApJ...964...36R,2024ApJ...977..173C,2025A&A...693A.278V}, and particularly the large programmes MINDS and JDISCS \citep{2024PASP..136e4302H,2025AJ....170...67A}.
Flux ratio measurements supported by astrochemical modelling can be used to determine the C/O ratio of the inner disc \citep{2018A&A...618A..57W,2021ApJ...909...55A}, which can have diverse compositions. Typically, inner discs around solar-mass T~Tauri stars are bright in water emission, while discs around low-mass stars show hydrocarbon emission indicating high C/O ratios \citep{2025A&A...702A.126G,2025A&A...699A.194A}, which suggests a trend across stellar masses. Solar-mass exceptions from this trend include, e.g., the carbon-rich DoAr~33 \citep[][]{2024ApJ...977..173C}, which may be explained by an unusually low accretion rate and the water-poor MY~Lup \citep{2025AJ....169..184S}, which possibly results from its high inclination. Alternatively, these cases could indicate unusual chemical or physical conditions. The water emission appears to be anti-correlated with the disc size and the presence of substructures in the outer disc \citep{2023ApJ...957L..22B,2025A&A...694A.147G}.

The main interpretation of the variety of the C/O ratios observed in inner discs is that this is a result of dust radial drift \citep{2011ApJ...743..147N,2013ApJ...766..134N,2020ApJ...903..124B,2023ApJ...957L..22B,2025ApJ...990L..72K}. Dust evolution and drift models predict water-rich inner regions in compact discs with uninterrupted pebble drift, whereas dust traps of various origins can intercept the drift, resulting in a lower water content in the inner disc and larger disc sizes \citep{2021ApJ...921...84K,2023ApJ...954...66K,2024A&A...686L..17M,2024A&A...691A..32N,2024A&A...691A..72L,2024ApJ...977...21E}. 
The observed trend with stellar mass can also be attributed to the effect of the drift being sensitive to the thermal structure: \citet{2023A&A...677L...7M} show that low-mass stars have faster carbon enrichment due to closer in snowlines. However, not all compact discs follow the hypothesis that their cold \ce{H2O} reservoir is enhanced following efficient radial drift \citep{2025A&A...699A.134T}. This suggests that some additional processes are at play, affecting the C/O balance in the inner disc. This could be the effect of dust \textit{leaking} from the trap \citep{2023A&A...670L...5S,2024A&A...686A.135P,2025ApJ...988...94H}, or possibly an effect of chemical evolution. Recent modelling of dynamically evolving discs with a more sophisticated surface chemistry model by \citet{2025A&A...701A.194P} showed that chemical processes can significantly change the disc composition at timescales longer than 2\,Myr and should be considered simultaneously with dust evolution, and highlighted the importance of initial composition and global ionisation rate. The ionisation rate and stellar mass can also be key factors in carbon enrichment of inner discs, particularly in the presence of dust traps \citep{2025A&A...701A.239S}.

Chemical reactions change the composition of the disc at timescales of $\sim1$\,Myr \citep{2018A&A...613A..14E}, slower than typical timescales of transport, particularly at higher levels of turbulence \citep[$\alpha>10^{-3}$;][]{2019MNRAS.487.3998B}, which supports the applicability of the freeze-out based models. Grain-surface reactions do typically operate at timescales $>1$\,Myr \citep{2018A&A...613A..14E,2018A&A...618A.182B,2025A&A...701A.194P}. However, cosmic-ray induced processes, particularly destruction of gas-phase methane and CO and their conversion to less volatile species can proceed at shorter timescales of a few 100\,kyrs \citep{2018A&A...613A..14E,2020ApJ...890..154P}. Cosmic ray ionisation is the main driver of chemistry in the midplanes of protoplanetary discs \citep{1999ApJ...519..705A}, because it triggers reactions in the disc midplane and is eventually responsible for the processing of the observable molecules \citep{2013ApJ...772....5C,2014ApJ...794..123C,2015ApJ...799..204C,2018ApJ...856...85S,2021ApJS..257...13A,2022ApJ...926L...2T,2022ApJ...932....6V}. At the same time methane is a key species necessary to obtain carbon-rich gas in discs, being the only main carbon carrier in the disc midplane when $\mathrm{C/O}>1$, and its initial abundance determines how high the C/O can become  \citep{2023A&A...677L...7M,2025A&A...699A.227H,2025A&A...701A.239S}. Therefore, its chemical destruction can significantly alter the carbon enrichment of the inner disc. The sensitivity of CO and \ce{CH4} chemical processing to the ionisation rate can also affect the observed inner disc C/O ratio \citep{2025A&A...701A.239S}.

In this paper, we investigate the chemical composition of a protoplanetary disc midplane using a model that combines dust evolution and transport with sophisticated gas-phase and surface chemistry.
Many of the observed protoplanetary discs are older than 1\,Myr~\citep[e.g.,][]{2023ASPC..534..539M}, thus their molecular compositions are likely affected by chemical processes. As discussed above, ice-phase reactions and cosmic ray driven chemistry govern the abundances of main carbon and oxygen bearing species at Myr timescales \citep{2016A&A...595A..83E,2018A&A...618A.182B,2019MNRAS.487.3998B}. Our goal is to analyse the combined effect of chemical reactions and dust radial drift on the composition of the inner regions of protoplanetary discs at timescales up to 10\,Myr.

The paper is structured as follows.
We describe the model in Section~\ref{sec:methods} before separately considering the effects of chemistry in Section~\ref{sec:no_transport} and dust drift in Section~\ref{sec:effect_of_transport}. Next, we consider their combined effect on the inner disc composition in Section~\ref{sec:baseline_chamistry}, in discs with and without dust traps (Sections~\ref{sec:c2o_in_discs_with_traps} and~\ref{sec:c2o_in_no_trap_discs}, respectively). After that, we consider the effect of the trap location in~\ref{sec:trap_location} and of the ionisation rate in Section~\ref{sec:ionisation_rate}. Finally, we discuss the model caveats and observational implications in Section~\ref{sec:discussion} and summarise our findings in Section~\ref{sec:conclusions}.

\section{Methods}
\label{sec:methods}

In the present work we simulate chemical evolution together with dust growth and drift in a viscously evolving disc. We use an astrochemical model with comprehensive treatment of surface chemistry by \citet{2015A&A...582A..88W} coupled to the model of viscous disc evolution adopting a two-population approximation of dust growth as done in \citet{2019MNRAS.487.3998B}. Compared to \citet{2019MNRAS.487.3998B}, the network by \citet{2015A&A...582A..88W} features more elaborate surface chemistry including two-body surface reactions, photodissociation of ices and reactive desorption. These reaction types are necessary to account for long-term chemical processes on dust grains at Myr timescales. This is the same chemical network also used in \citet{2018A&A...618A.182B}.
We explore different efficiencies of radial drift, consider models with and without dust traps, and investigate the effect of chemical reactions by varying incident cosmic ray ionisation rate and the contribution from cosmic ray induced dissociation of ices.

\subsection{Gas and dust evolution}
\label{sec:model_viscous}

We adopt the viscous evolution code by \citet{2017MNRAS.469.3994B} that solves the equation by \citet{1974MNRAS.168..603L} for multi-fluid gas consisting of multiple gas and dust components. The code uses the classic $\alpha$ parametrisation of viscosity $\nu = \alpha c_{\rm s} H$, with sound speed, $c_{\rm s}$, and local disc scale height, $H$ \citep{1973A&A....24..337S}. In this work we use a default value of $\alpha=10^{-3}$. Initially, the surface density radial profile is a tapered power law with an exponent of $-1$, a characteristic radius of 100\,au, and has an initial gas mass of $0.01$\,$M_{\odot}$. The central star has solar mass, an effective temperature of $4000$\,K and a radius of $2.5$\,$R_{\odot}$. The midplane temperature $T(R)$ is determined by a combination of viscous and irradiative heating following the prescription by \citet{2019A&A...632A...7L}.

Dust evolution is described following the two-population model \texttt{two-pop-py} by \citet{2012A&A...539A.148B}. The initial mass fraction of refractory grain cores (including the refractory carbon component, and excluding ice mantles) is 0.0056 relative to the gas. This mass fraction is determined by the adopted chemical composition (see Section~\ref{sec:chemical_model}). The combined mass of dust and ice relative to the gas used in the grain growth calculations is 0.012, close to the standard ISM value of 0.01. The two populations of grains have different dynamics: small grains have fixed size and are dynamically coupled to the gas, while larger grains can grow and  move relative to gas, i.e., drift toward pressure maxima. The model analytically approximates collisional growth of dust grains to their maximum size where growth is restricted by fragmentation or radial drift, depending on local conditions. The ices frozen on dust are assumed to be distributed such that all of the grains have the same ice mass fraction.

Fragmentation is characterised by the fragmentation velocity $v_{\rm frag}$, which in most of our models equals either 10 or 1\,m~s$^{-1}$ for icy and non-icy grains, respectively. This value corresponds to the dust grains being resilient to collisional fragmentation in most of the disc, thus growing to larger sizes and efficiently drifting though the disc (fast transport). Combined with the relatively high value of $\alpha=10^{-3}$, such resilient grains allow to reproduce dust emission in multi-wavelength ALMA observations, as shown by \citet{2026MNRAS.545f2128T} using MAPS sources \citep{2021ApJS..257....1O,2021ApJS..257...14S}, although the AGE-PRO data suggests that low turbulence and low fragmentation velocities might be a better match for the dust emission \citep{2026MNRAS.548ag423L}. Laboratory experiments and observational data tend to favour lower values of $v_{\rm frag}\approx 1$\,m~s$^{-1}$  \citep{2018MNRAS.479.1273G,2019ApJ...873...58M,2024A&A...682A..32J,2024NatAs...8.1148U}. Using $v_{\rm frag}=1$\,m~s$^{-1}$ in combination with weaker turbulence ($\alpha=10^{-4}$) would produce similar drift velocities, so we choose to use high values of $\alpha=10^{-3}$ and $v_{\rm frag}=10$\,m~s$^{-1}$ and explore drift efficiency by varying $v_{\rm frag}$. To do that, 
we consider a model with $v_{\rm frag}=1$\,m~s$^{-1}$ regardless of the presence of ice mantles (slow transport), to analyse the effect of chemistry in models with weak dust transport.

To add a dust trap to the disc, we introduce a planet of $1$\,$M_{\rm Jup}$ at a fixed radial distance. This planet creates a perturbation in the gas surface density, modelled according to the prescription by \citet{2020ApJ...889...16D}. The gap in the gas distribution results in a pressure maximum exterior to the planetary orbit, where dust grains become trapped. For a planet at 10\,au, the dust trap is formed at $\approx15$\,au distance for the adopted model parameters. The planet itself does not evolve in any way, and its mass and location are solely employed to modify the radial profile of $\alpha$ multiplying it by a gap profile factor \citep[see][for details]{2020ApJ...889...16D}. The modified $\alpha$ is used to calculate viscosity, locally changing gas surface density and creating a pressure bump where dust can be trapped. The diffusion rate and collision velocities between dust grains that regulate the maximum dust size also depend on $\alpha$; for them the unchanged value of $\alpha=10^{-3}$ is used.

The adopted two-population dust evolution model is an analytical estimate of full dust evolution that was developed and optimised for smooth discs \citep{2012A&A...539A.148B}. Nevertheless, it has also been used for modelling discs with substructures \citep{2023ApJ...954...66K,2024ApJ...977...21E,2025A&A...694A..79S,2025A&A...701A.239S}. In their code comparison, \citet{2026arXiv260322550E} recently showed that \texttt{two-pop-py} depletes the dust reservoir faster and traps dust grains more efficiently than more sophisticated models. Another study by \citet{2026arXiv260411925H} demonstrates that dust traps are leakier, i.e. less efficient in blocking dust transport, than previously thought. To properly capture dust evolution and transport in the vicinity of a planet-induced gap would require a better approximation of dust fragmentation and coagulation and 3D hydrodynamic modelling  \citep{2022ApJ...935...35S,2025ApJ...994L..44V}. We acknowledge that the adopted treatment of dust evolution has caveats but proceed to use it as an optimal option for combining with computationally heavy chemical simulations. We discuss these caveats in Section~\ref{sec:discussion_trap_chemistry}.

We run the models up to a time of 10\,Myr. At this timescale, the disc evolution should also be affected by photoevaporation.
For simplicity, in this work we neglect its contribution.

\subsection{Chemical model}
\label{sec:chemical_model}

The chemical network in the \citet{2019MNRAS.487.3998B} model already included a relatively large chemical network, but with only hydrogenation reactions on dust grain surface. We updated it to the model by~\citet{2015A&A...582A..88W}, which was also described by \citet{2016A&A...595A..83E}. The network is based on the Rate12 chemical network \citep[][]{2013A&A...550A..36M} and contains more complex grain surface chemistry by \citet{2008ApJ...682..283G}. The network includes gas-phase two-body reactions, photodissociation and photoionisation, ionisation by cosmic-rays and cosmic-ray-induced photons and radioactive nuclides. Transitions between the gas and the ice phases are described by adsorption to grain surface, thermal desorption, photo-desorption by cosmic-ray induced photons, and reactive desorption of products of surface chemistry. Reactions on dust grains are described by the Langmuir-Hinshelwood mechanism, with \ce{H} and \ce{H2} quantum tunnelling between binding sites and reaction barriers \citep[see][for more details]{2015A&A...582A..88W}. 

Because we focus on the disc midplane, photo-reactions by UV irradiation, X-ray ionisation, Ly-$\alpha$ photons, and reactions with excited \ce{H2} are not considered. We note that the ionisation by X-ray induced secondary photons can be higher than CR ionisation even in the midplane, e.g. if the emission source is hotter than $3 \times 10^7$\,K \citep[see, e.g., Figure~13 in][]{2009ApJS..183..179B}, but here we assume that this is not the case and neglect its contribution. Ionisation is provided by cosmic ray-induced photons (CRPHOT). The rates of the CRPHOT reactions in the model are calculated following \citet{1989ApJ...347..289G}, where interstellar dust size distribution and classic 0.01 dust-to-gas mass ratio are adopted, not adjusted to the protoplanetary disc midplane conditions.
Cosmic ray attenuation is included following the simpler approximation by \citet{2000ApJ...543..486S}, with the attenuation column of 96\,g~cm$^{-2}$. The default value of incident CR ionisation rate is $\zeta=10^{-17}$\,s$^{-1}$, which we decrease or increase by an order of magnitude in some models. We calculate the column in the vertical direction only and include a factor of $0.5$ to account for the CR flux from both sides of the disc. We also add a residual ionisation rate by radioactive nuclides with $\zeta_{\rm R}=6.5 \times 10^{-19}$\,s$^{-1}$ \citep{2004A&A...417...93S}. 

Surface chemistry is one of the key properties of the updated model. We adopt the efficiency of reactive desorption equal to 10\,\%, and the ratio between diffusion and binding energy equal to 0.3 \citep{2014A&A...569A.107K,2014JChPh.141a4304M,2016A&A...585A..24M}. The surface area of dust grains is calculated as the mean surface area of the dust ensemble with local minimum and maximum sizes and dust mass fraction, assuming the power law exponent of $-3.5$. The model treats surface chemistry in the two-phase approach, without distinguishing between surface layers and the bulk of ice mantles, unlike some other chemical models of molecular clouds and protoplanetary discs \citep[see, e.g.,][]{1993MNRAS.263..589H,2013ApJ...765...60G,2019ApJ...885..146R,2022ARep...66..393B}. Instead, we assume that grain surface reactions are active only in the upper two monolayers of the ice mantles, by applying an effective reduction factor to the rates of surface two-body reactions and desorption reactions. However, no such factor is applied to the ice-phase reactions with cosmic ray-induced photons, as high energy photons are assumed to easily pass through ice and induce reactions in the whole ice mantle.

High-energy cosmic ray induced photons can dissociate species inside the bulk of ice mantles, as well as on the grain surface and provide surface radicals responsible for the formation of complex molecules on dust grains \citep{2009A&A...496..281O,2009A&A...504..891O}. In astrochemical modelling, the rates of CRPHOT reactions for ice-phase species are assumed to be the same as for the gas phase, implying that the whole of the ice mantle is affected by the process \citep[e.g.,][]{2008ApJ...682..283G}. It can potentially lead to overestimation of their role in surface chemistry, as the products of these reactions in the model can subsequently participate in reactions on the surface, when otherwise they should be locked in the bulk. In contrast with the surface where species can move freely, the reaction products in the bulk diffuse very slowly and might only be able to react with the immediately nearby molecules. Our \textit{baseline chemistry} models do not include the ice-phase reactions with cosmic-ray induced photons. We explore their effect with the \textit{full chemistry} models and vary their contribution in Section~\ref{sec:ionisation_rate}.

In the dynamic disc model, gas- and ice-phase species are transported with the gas and dust components, respectively. Contribution from ice mantles is included in both size and mass of dust grains, assuming an average density of solids of 1\,g~cm$^{-3}$. The chemical evolution is a computationally demanding problem, so we only solve kinetic equation every few transport steps. The time step for chemistry is $\sim100$\,yr in all models, which corresponds to 5 dynamic steps in the smooth disc models. In the regions around the outer disc edge, where the dust-to-gas mass ratio and temperature are very low ($<10^{-6}$ and $<15$\,K respectively), the system of rate equations becomes harder to solve. To avoid computational problems, we only consider adsorption and desorption of volatiles in these regions because the mobility of species will be slow, and the surface area available for surface reactions is low. This region contains $<1\%$ total gas mass before 3\,Myrs. At times $>8$\,Myr, this fraction grows above $10\%$ due to viscous spreading of gas and dust depletion by radial drift. As the viscous spreading is directed outwards in these region, its contribution to the disc chemical composition inside 100\,au is negligible.

The initial chemical composition is adopted from \citet{2023A&A...677L...7M}. It is based on the Solar elemental abundances from \citet{2009ARA&A..47..481A} and assumes that the volatiles are initially in the ice phase in form of simple molecular ices: CO, \ce{N2}, \ce{CH4}, \ce{CO2}, \ce{NH3}, \ce{H2S}, and \ce{H2O}. The rest of the elements, including refractory carbon, is assumed to comprise the cores of dust grains, resulting in a dust-to-gas mass ratio of 0.0056 without ices and 0.012 with ices. This setting implies that the disc retains the composition of the parent molecular cloud, which is supported by the detection of methanol in discs around Herbig~Ae/Be stars where its formation should be suppressed due to higher temperatures $\gtrsim 20$\,K because it forms from CO ice \citep{2021NatAs...5..684B, 2025ApJ...982...62E}. The adopted composition from \citet{2023A&A...677L...7M} has been debated as having too much \ce{CH4}, which could affect the degree of chemical enrichment \citep{2025A&A...699A.227H}. We discuss the caveats of this assumption in Section~\ref{sec:high_c2o_where}.

The C/O ratio in the inner disc can also be affected by the irreversible destruction of the carbonaceous component of dust at higher temperatures due to oxydation, pyrolisis or other chemical processes \citep{2010ApJ...710L..21L,2017A&A...606A..16G,2019ApJ...870..129W}. This process can release additional carbon to the gas, which then can diffuse with the gas and help sustain more C in the disc, saving it from accretion to the star with solid grains and increasing the gas-phase C/O ratio \citep{2021SciA....7.3632L,2024MNRAS.535..171P,2025A&A...699A.227H}. To consider this process in our model, we follow \citet{2025A&A...699A.227H} and add decomposition of refractory carbon at $T>350$\,K. Solid C, which constitutes around 25\,\% of solid refractory grain material (the rest being silicates and metals), is turned into acetylene \ce{C2H2}, which can then react with other species in the gas phase, and diffuse with the gas. We include this process in some of the models to constrain its effect on the carbon and oxygen balance in the inner disc.

\subsection{Models considered}
\label{sec:set_of_models}

Chemical reactions and dust transport are the main processes redistributing carbon and oxygen in the disc and affecting the C/O ratios. We consider a set of models that covers different chemistry and drift regimes to disentangle these effects, and their properties are summarised in Table~\ref{tab:models}.

\begin{table}
    \centering
    \begin{tabular}{cccc}
        \hline
    Chemistry & Transport & Dust trap & CR ionisation \\
        \hline
        baseline / full & no & no & $\zeta = 10^{-17}$ \\
        freeze-out only & fast &  no / at 15\,au & $\zeta = 10^{-17}$\\
        baseline / full   & slow &  no  & $\zeta = 10^{-17}$  \\
        baseline & fast &  no / at 15\,au  & $\zeta = 10^{-18}$ / $10^{-17}$ / $10^{-16}$ \\
        full   & fast &  at 15\,au  & $\zeta = 10^{-18}$ / $10^{-17}$ / $10^{-16}$\\
        full   & fast &  at 15\,au  & $\zeta = 10^{-17}$,  \\
           &  &  & $C_{\rm iCR} =10^{-1}$ / $10^{-2}$ / $10^{-3}$ \\
        baseline / full  & fast &  at 7, 45, 100\,au  & $\zeta = 10^{-17}$\\
        baseline C & fast &  no / at 15\,au & $\zeta = 10^{-17}$ \\
        \hline
    \end{tabular}
    \caption{Considered model parameters. Transport regimes include models with no gas or dust transport (no), with gas and dust transport and resilient icy grains ($v_{\rm frag} = 10$\,m~s$^{-1}$, fast), and with fragile icy grains ($v_{\rm frag} = 1$\,m~s$^{-1}$, slow). The dust trap is either absent, or set at a specified radial distance. $\zeta$ (s$^{-1}$) is the incident cosmic ray ionisation rate,  $C_{\rm iCR}$ is the efficiency factor for the cosmic-ray induced dissociation of ices (see Section~\ref{sec:CRPHOT_ices_trap}).}
    \label{tab:models}
\end{table}

To remove the dynamical effects entirely, in Section~\ref{sec:no_transport} we consider the starting models with no gas or dust transport (but with dust growth) similar to the models by \citet{2016A&A...595A..83E}. Then we move on to our main set of models that have different transport regimes. Models with a high fragmentation velocity for ice-covered grains ($v_{\rm frag}=10$\,m~s$^{-1}$) have large grains and efficient dust drift (\textit{fast transport} models). To restrict dust transport, we set $v_{\rm frag}=1$\,m~s$^{-1}$ for both icy and bare dust grains; this decreases the maximum dust size and slows down the radial drift (\textit{slow transport}). These models are considered in Section~\ref{sec:effect_of_transport}. Another way to restrict dust transport is to add a dust trap that will stop large grains from drifting all the way to the star. We include a dust trap at 15\,au created by a Jupiter-mass planet at 10\,au to our \textit{fast transport} models, and vary the planet location in Section~\ref{sec:trap_location}.
 
To assess the effect of chemistry, we consider the \textit{freeze-out only} models, where no chemistry is active apart from adsorption and thermal desorption. These models serve to distinguish the effects of dust evolution and transport from the effects of chemistry and to validate our results against the predictions of other models. The effect of chemistry is studied based on our \textit{baseline chemistry} models with no ice-phase CRPHOT reactions. The contribution of these reactions is varied in  Section~\ref{sec:ionisation_rate}, and they are fully included in \textit{full chemistry} models that serve as an extreme case, where the effect is most strongly manifested and likely overestimated. From a physics point of view, this case corresponds to highly porous and fluffy dust grains, that have large surface area-to-mass ratio \citep{2009ApJ...707.1247O} and therefore a larger fraction of mantle participating in surface reactions. The presence of such porous grains have been inferred in protoplanetary discs by polarised scattered light observations \citep{2019ApJ...885...52T,2023ApJ...944L..43T}. We vary the incident CR ionisation rate by a factor of ten ($10^{-18}$ and $10^{-16}$\,s$^{-1}$) to control the contribution of chemistry. Additionally, we include carbon grain destruction in our \textit{baseline C} models and analyse its effect in Section~\ref{sec:carbon_destruction}.

\section{Results}
\label{sec:results}

\subsection{Chemistry in discs with no radial transport}
\label{sec:no_transport}

\begin{figure*}
\includegraphics[width=0.66\columnwidth]{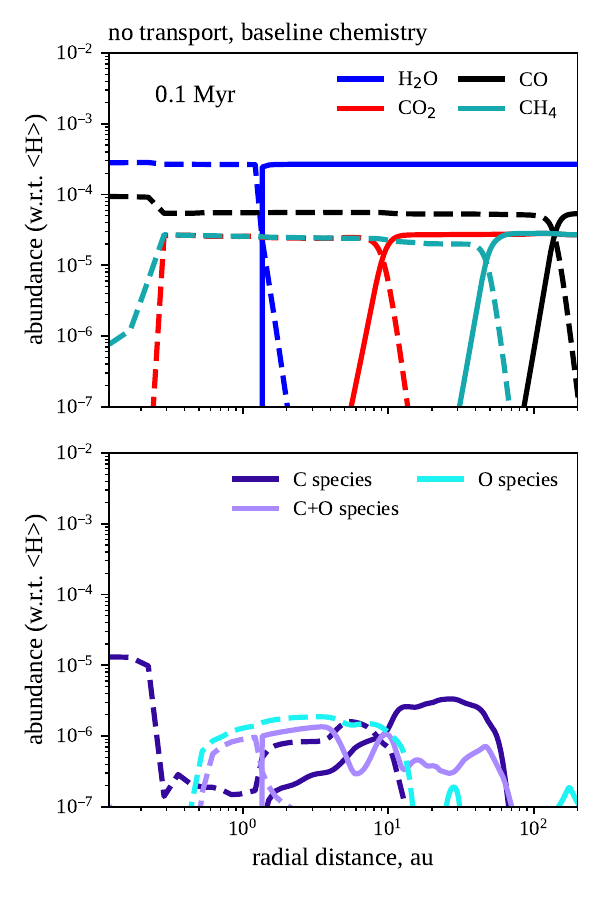}
\includegraphics[width=0.66\columnwidth]{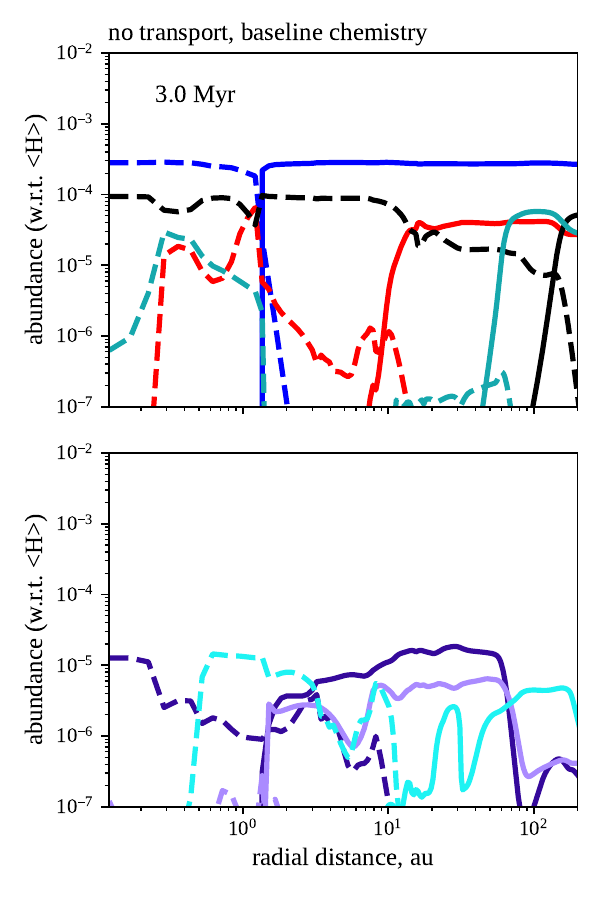}
\includegraphics[width=0.66\columnwidth]{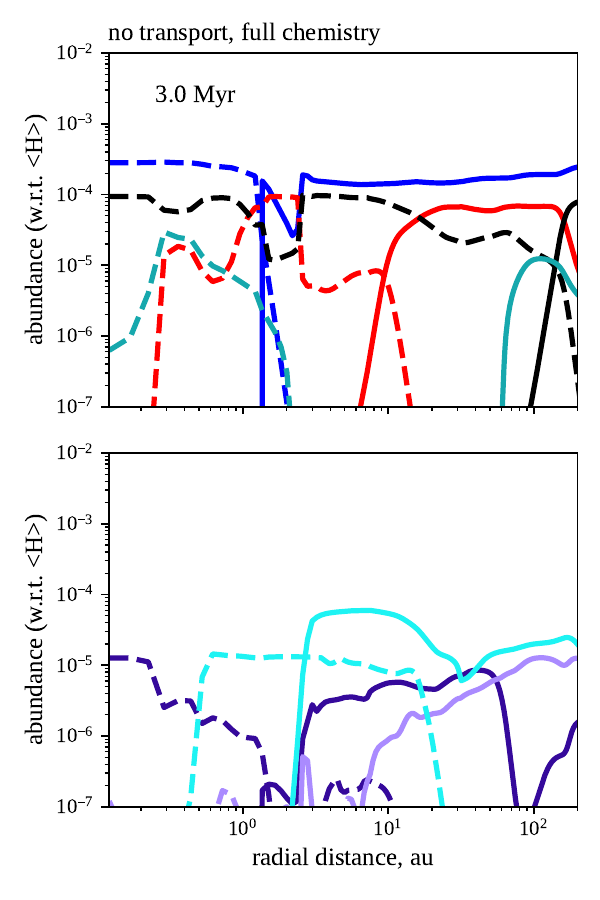}
    \caption{Abundances (w.r.t. H nuclei) of main C and O carriers in the baseline and full chemistry models without transport. \textit{C species} represent the combination of all species containing carbon, but not oxygen; \textit{O species} represent the species containing oxygen, but not carbon; \textit{C+O species} represent the  species containing both carbon and oxygen. From all the grouped species, the initially present \ce{H2O}, \ce{CO2}, CO and \ce{CH4} are excluded. Solid lines represent ice-phase species, and dashed lines represent gas-phase species.}
    \label{fig:no_transport_radial}
\end{figure*}

In the absence of gas or dust transport, the only way to change gas- and ice-phase elemental abundances is provided by chemistry, including freeze-out and desorption. We will first consider how the abundances of main volatile species evolve due to chemistry in a model with no viscous transport or dust radial drift. In this purely chemical model with no transport we still account for the collisional growth of dust grains; they reach a maximum size of $\sim0.1-1$\,cm within $10^{4}-10^{5}$\,yrs in most of the disc.

Figure~\ref{fig:no_transport_radial} shows radial distributions of main C- and O-bearing species in the models with chemistry and no transport. Starting with only \ce{H2O}, \ce{CO2}, CO and \ce{CH4}, by 0.1\,Myr chemistry already produces a noticeable amount of new species containing carbon and oxygen. This is due to cosmic ray ionisation triggering destruction of CO and methane at distances of up to tens of au.
CO is mainly dissociated by \ce{He^+} ions produced by CRPHOT ionisation, creating \ce{C^+} and \ce{O}. Methane is destroyed by \ce{C^+}, as well as by \ce{He^+}. Both \ce{CO} and \ce{CH_4} additionally react with a suite of other ions (\ce{H3+}, \ce{H+}, \ce{N2H+} etc.), and produce more complex ions (e.g. \ce{HCO+}, \ce{C2H2+}, \ce{C2H3+}, \ce{CH5+}, \ce{CH2+}, and \ce{CH3+}). These ions continue to react and eventually either return carbon back to CO or \ce{CH_4}, or form more complex organic species, such as \ce{C2H2}, \ce{C2H4}, \ce{C2H6}, \ce{CH3CHCH2}, \ce{CH3CCH}, and \ce{CH2CCH2}. Since most of these new species have higher binding energies with snowlines in the range of $1-10$\,au, they immediately freeze out, transferring carbon from the gas to the ice phase. Overall, at $r>10$\,au, methane is destroyed an order of magnitude faster than \ce{CO}. \ce{CO_2} is also dissociated by CRPHOT and \ce{He^+}, releasing carbon and oxygen. Inside the \ce{CO2} snowline, the released atomic oxygen is processed in gas-phase reactions into \ce{SO} and \ce{SO2}, also producing small amounts of \ce{NO} and \ce{N2O}. The C+O species form mostly in the ice phase from the frozen out and processed products of the gas-phase cosmic ray ionisation; the most abundant of them are \ce{NH2CHO}, \ce{OCS} and \ce{H2CO}, with some \ce{CH2CO}, \ce{C3O} and \ce{HCOOH}. In the very inner disc, collisional dissociation and two-body reactions in the gas phase turn \ce{CO2} and \ce{CH4} into \ce{CO} and water, the most stable species at inner-disc temperatures and densities, as well as forming some \ce{H2CS} and \ce{HC3N}.

By 3\,Myr, the amount of newly formed species increases. Their abundances relative to H nuclei reach $10^{-6}-10^{-5}$ and  become comparable with those of the primordial species. Between the water and \ce{CO2} snowlines, \ce{CO2} is transformed into CO and complex organic molecules (COMs). Outside the \ce{CO2} snowline, CO turns into \ce{CO2} ice via ice-phase reactions, and outside the \ce{CH4} and CO snowlines, methanol forms on the dust surface \citep{2016A&A...595A..83E,2018A&A...613A..14E,2018A&A...618A.182B}. All other organics continue to form at $1-80$\,au, accumulating carbon and oxygen from the destruction of the main volatiles. 

By $3-10$\,Myr, the processing of the initial volatiles continues both in the ice and in the gas. In the baseline chemistry model, all gas-phase CO beyond 10\,au is converted to \ce{CO2}, \ce{CH4} and COMs. After $\approx7$\,Myr, no major volatile remains in the gas phase beyond the \ce{CO2} snowline, because they are all converted to more refractory species and frozen out on dust grains, reducing the metallicity of the gas by more than three orders of magnitude. Overall, the chemical evolution of the disc slows down after 3\,Myrs. However, the chemical processing before 3\,Myr is able to affect how radial drift redistributes elements across the disc.

Among these processes, those that channel elements into species with significantly different binding energies have the largest effect on the C/O, as they affect the delivery of these species to the inner disc regions. Carbon from volatile \ce{CH4} ($E_{\rm b}=1252$\,K, snowline at $\approx50$\,au) converted to less volatile COMs ($E_{\rm b}\sim 5000$\,K, snowlines at $\approx1-2$\,au) will be more strongly affected by the radial drift, and delivered to the inner disc much faster, together with water ice. As was previously shown by other authors \citep{2019MNRAS.487.3998B,2020ApJ...890..154P}, conversion of \ce{CH4} to less volatile hydrocarbons can lead to carbon enrichment of the inner disc due to dust radial drift. Similarly, conversion of CO to methanol moves the elements to less volatile COMs affected by dust transport, although keeping the same C/O ratio. Finally, conversion of oxygen from water ($E_{\rm b}=4880$\,K, snowline at $\approx1$\,au) to much more volatile \ce{O2} ($E_{\rm b}=898$\,K, snowline at $>100$\,au) in the full chemistry model will on the contrary make some fraction of oxygen less affected by dust transport. Formation of SO ($E_{\rm b}=1800$\,K, snowline at $\approx15$\,au) or \ce{SO2} ($E_{\rm b}=3010$\,K, snowline at $\approx5$\,au) should have a similar, but much weaker effect in relation to the transport.

In the full chemistry model, water ice is additionally being destroyed by CRPHOT dissociation, releasing oxygen in form of OH radicals on the dust surface. This leads to more efficient conversion of CO to \ce{CO2} ice and a higher abundance of methanol (C+O species at $>50$\,au on the lower right panel of Figure~\ref{fig:no_transport_radial}), a higher abundance of SO, and efficient formation of molecular oxygen and \ce{H2O2} on dust. After 3\,Myr, CO gas survives thanks to its additional production from CRPHOT dissociated water and organic ices. Water ice is depleted by a factor of $>3$, and its oxygen is recycled into \ce{CO2}, \ce{SO} and other molecules. Similar chemical evolution of the main volatiles was demonstrated by \citet{2016A&A...595A..83E} at 1\,Myr (see their Figures~4 and A.1). More recent modelling by \citet{2025A&A...701A.194P} using the same chemical network presented here demonstrates similar trends regarding \ce{CH4} depletion, conversion between CO and \ce{CO2}, and the formation of carbon-rich organics at 3\,Myr (see their Figure~3c). Note that the model of \citet{2025A&A...701A.194P} includes erosion of carbon grains inside 5\,au, which significantly affects the C/O ratio and chemistry in this region. We exclude the effect of carbon grain destruction from our main modelling and consider it in Section~\ref{sec:carbon_destruction}.

\subsection{Effect of dust transport on the evolution of the C/O ratios}
\label{sec:effect_of_transport}

In this section, we look at the C/O ratios throughout the whole disc and analyse their evolution due to the interplay between transport and chemistry. We focus on smooth discs without dust traps here.

Gas and dust are subject to viscous disc evolution and diffusion, and their efficiency is set by the value of $\alpha=10^{-3}$. Dust is additionally affected by radial drift, the efficiency of which depends on the maximum dust grain size. Dust size is also sensitive to $\alpha$ because that determines the collision velocity of dust grains, and to the fragmentation threshold $v_{\rm frag}$. Here we compare the evolution of the C/O ratios in the models with baseline chemistry and different transport regimes. We consider the model with no transport of dust or gas (presented in Section~\ref{sec:no_transport}), and models with both gas and dust transport, fast and slow, regulated by  $v_{\rm frag}$ (see Section~\ref{sec:set_of_models}).

\begin{figure*}
\includegraphics[width=0.97\columnwidth]{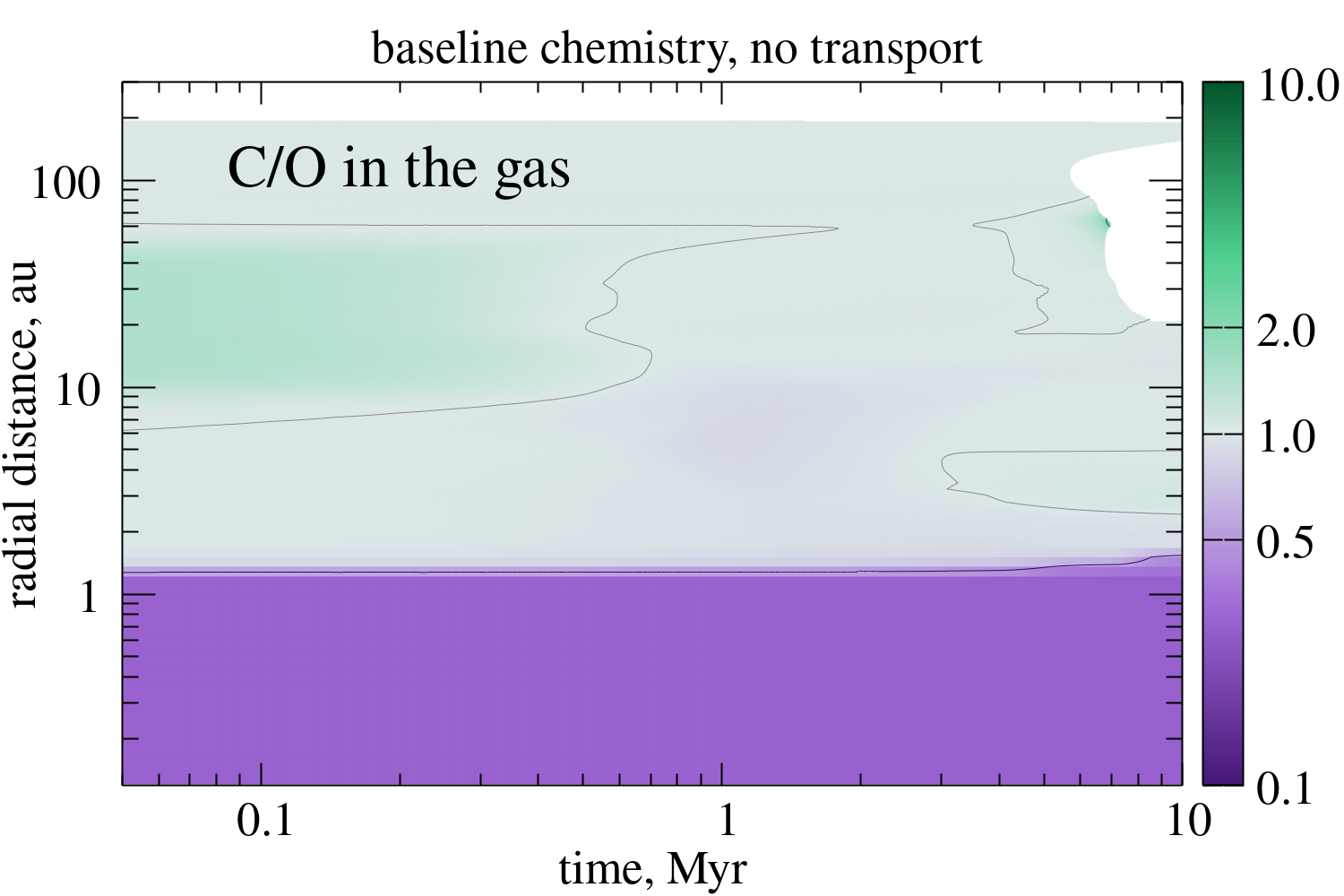}
\includegraphics[width=0.97\columnwidth]{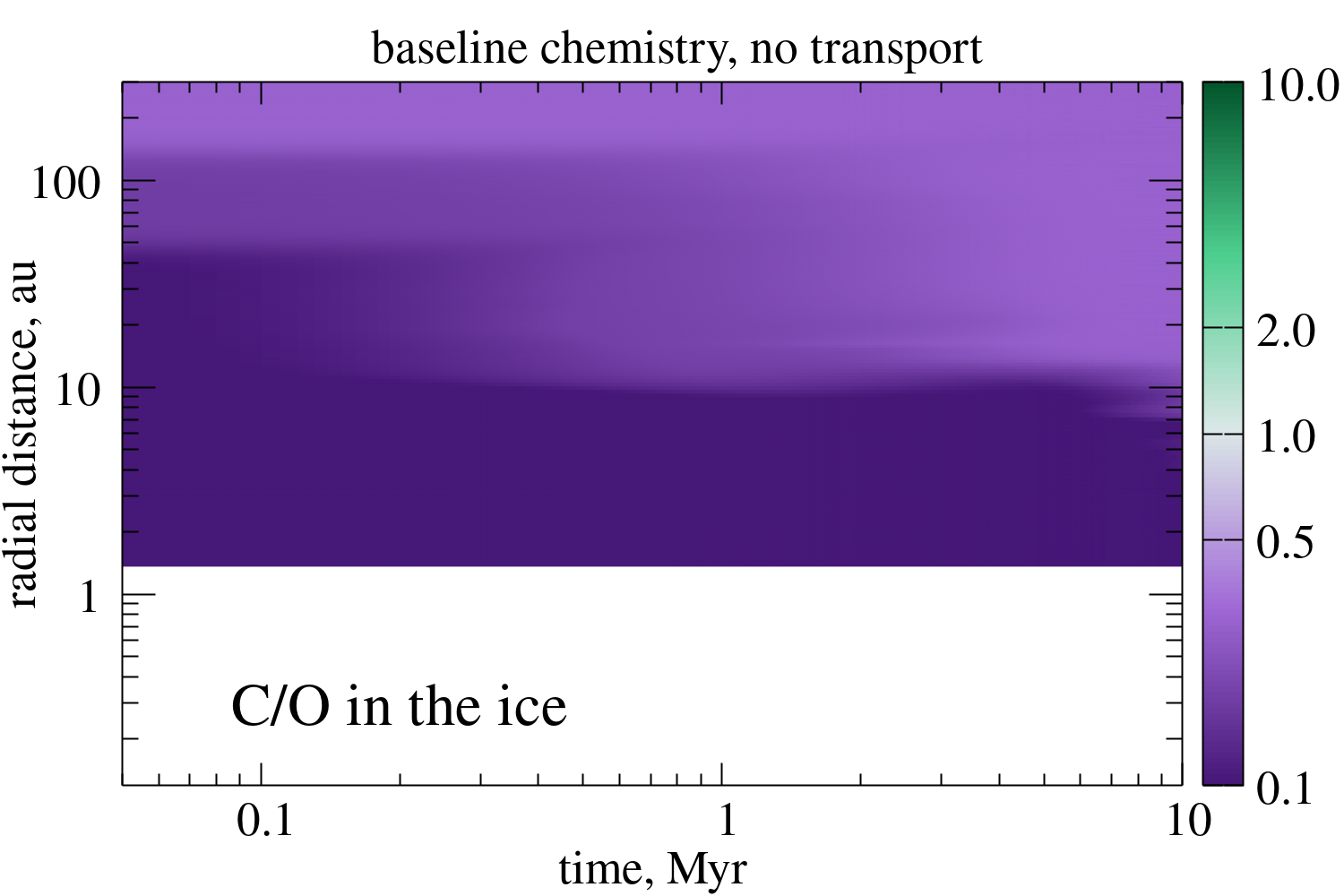}
\includegraphics[width=0.97\columnwidth]{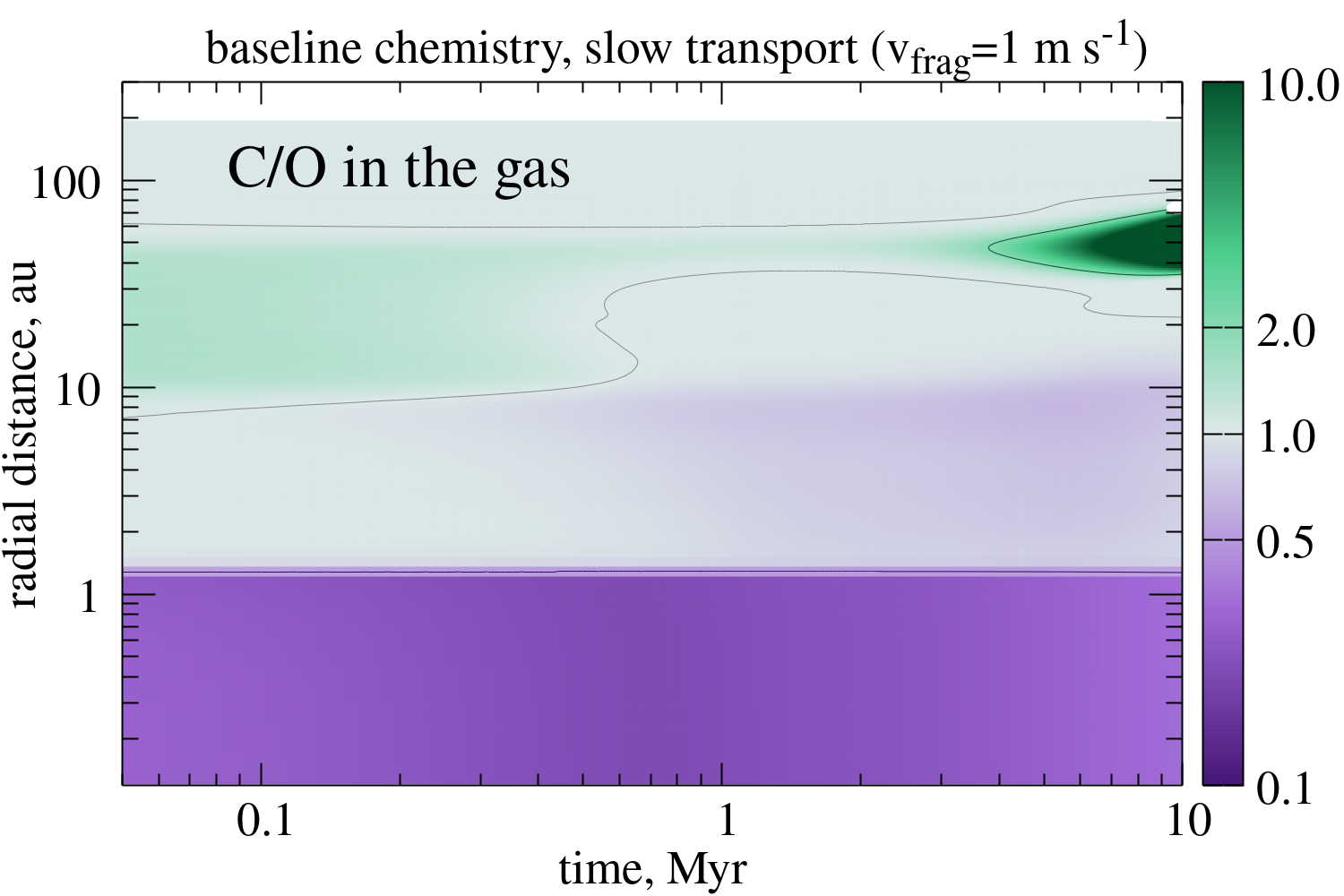}
\includegraphics[width=0.97\columnwidth]{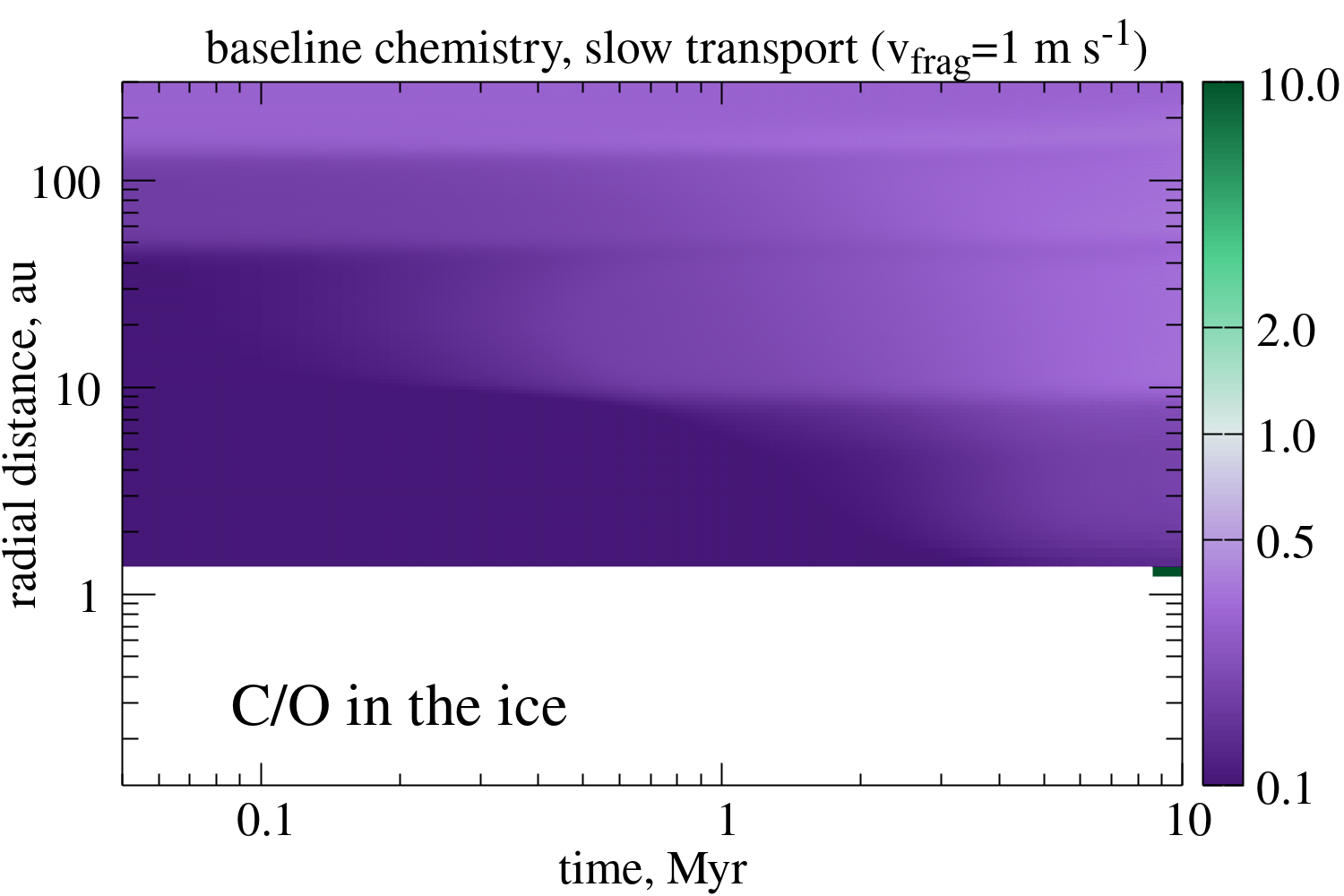}
\includegraphics[width=0.97\columnwidth]{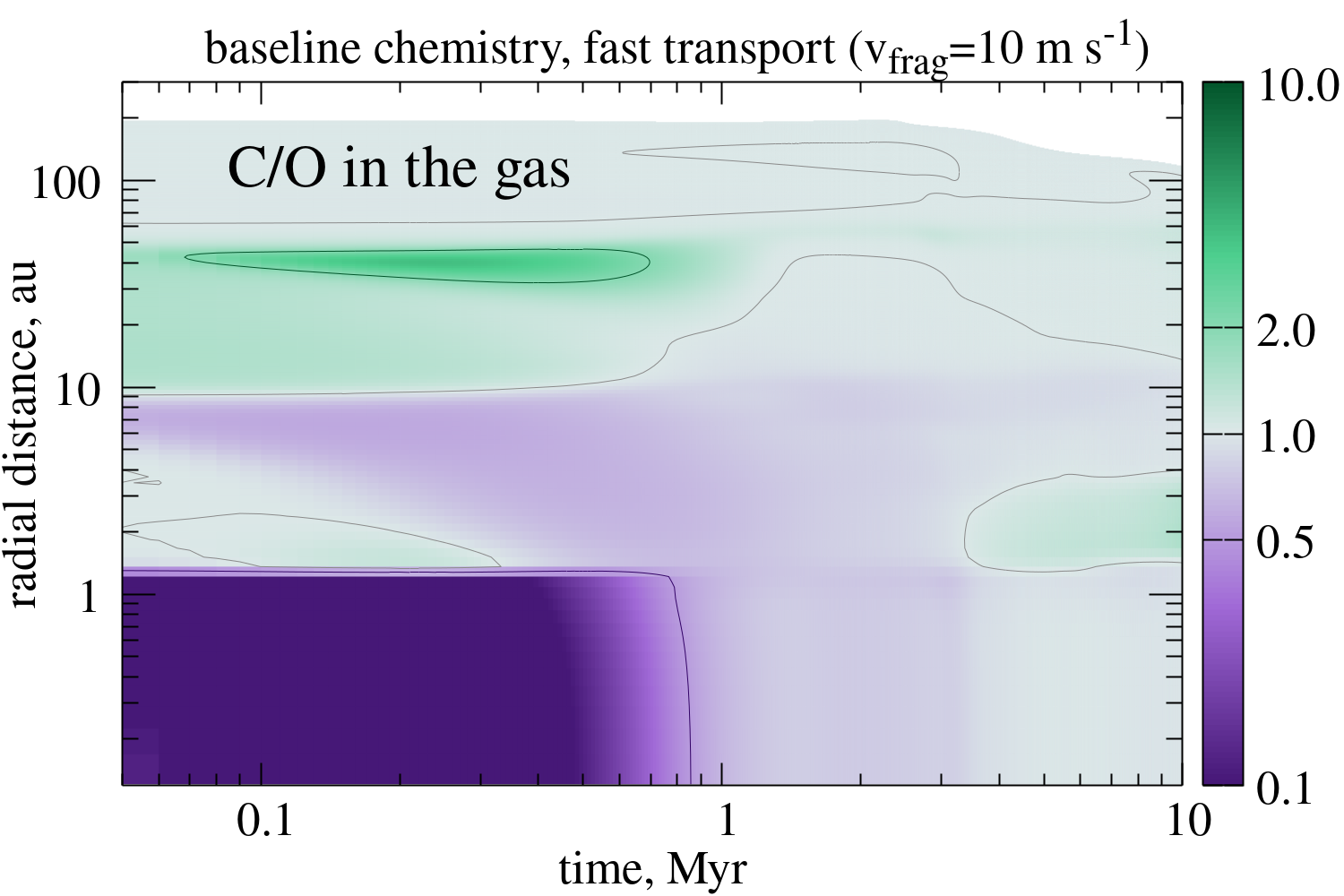}
\includegraphics[width=0.97\columnwidth]{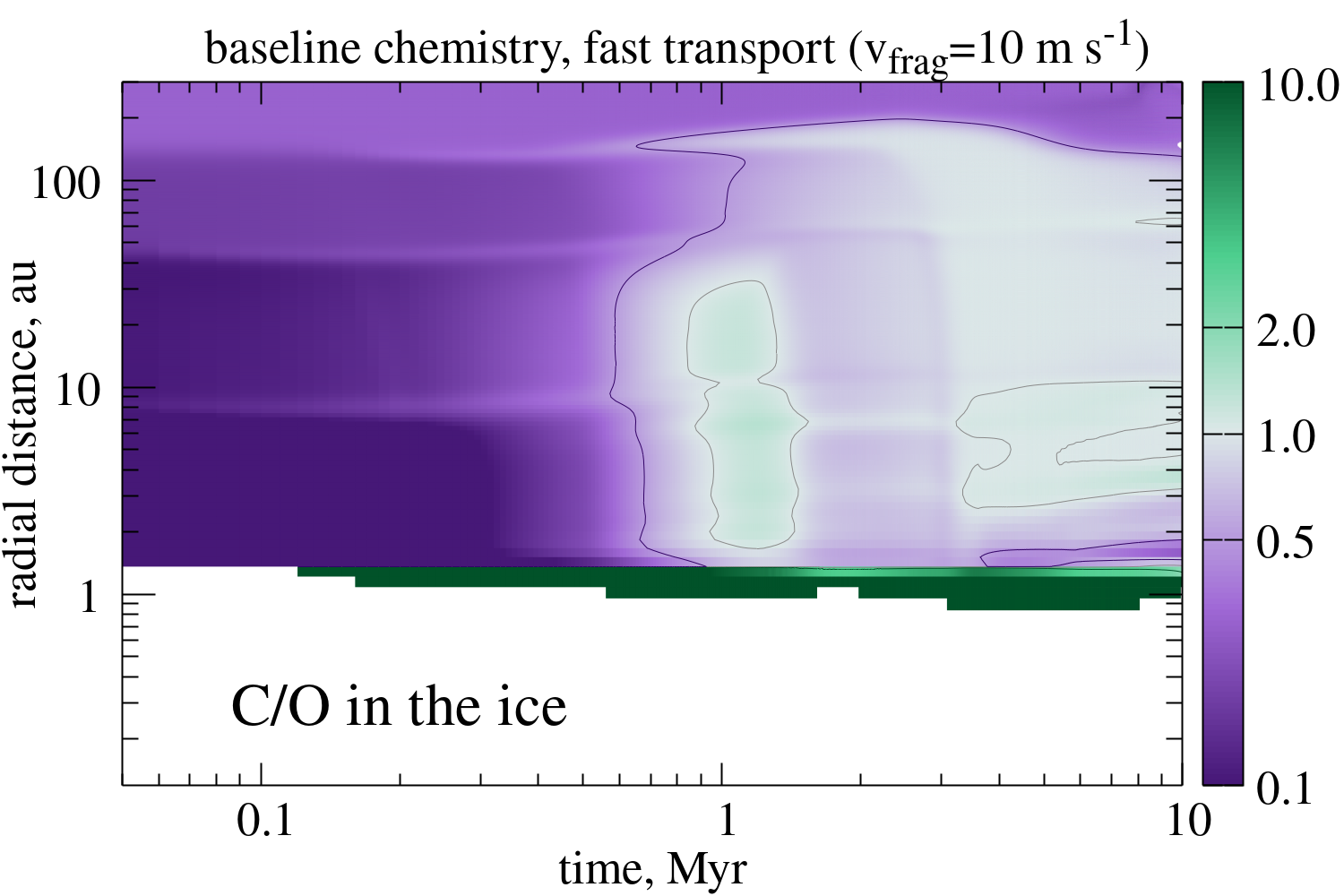}
\includegraphics[width=0.97\columnwidth]{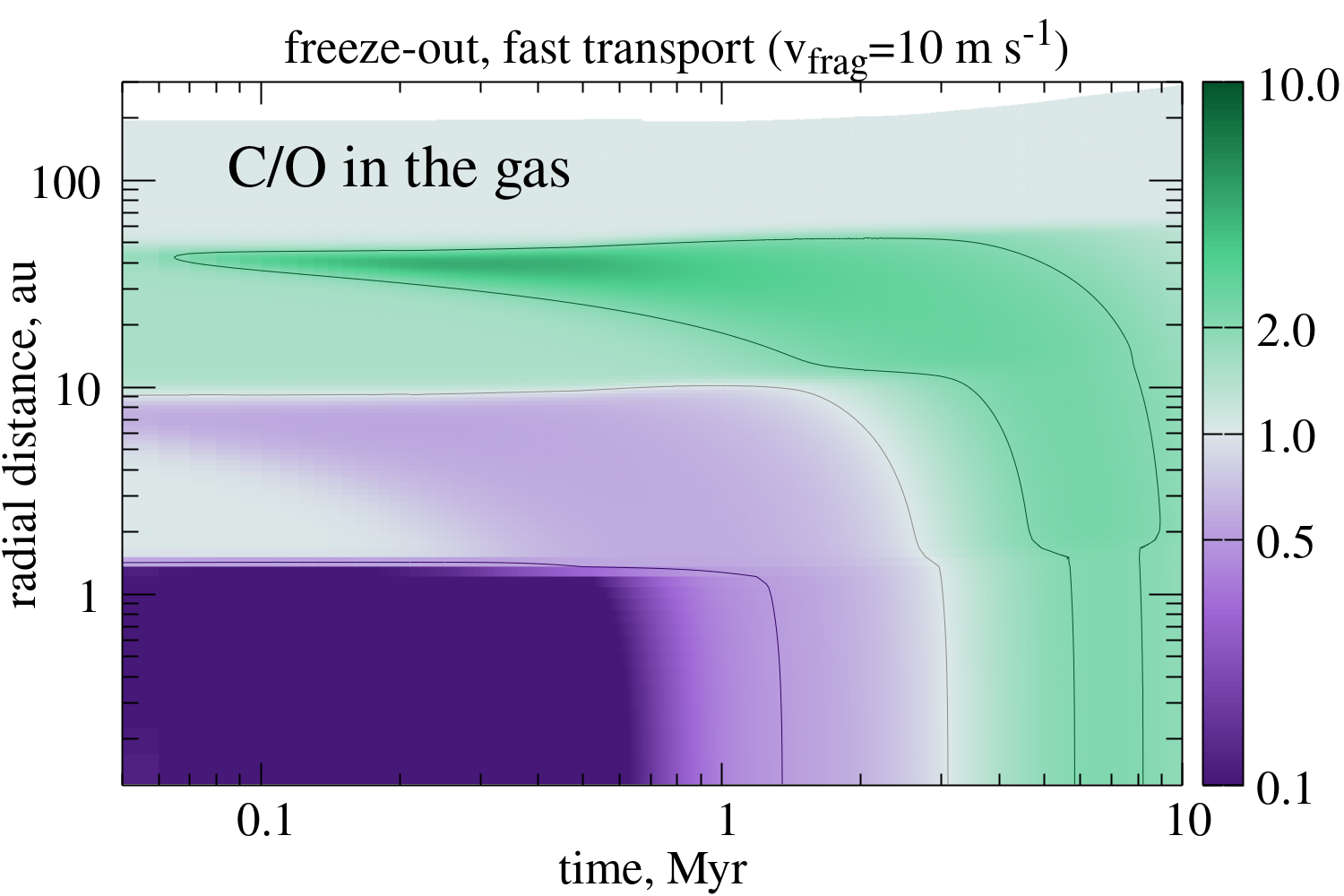}
\includegraphics[width=0.97\columnwidth]{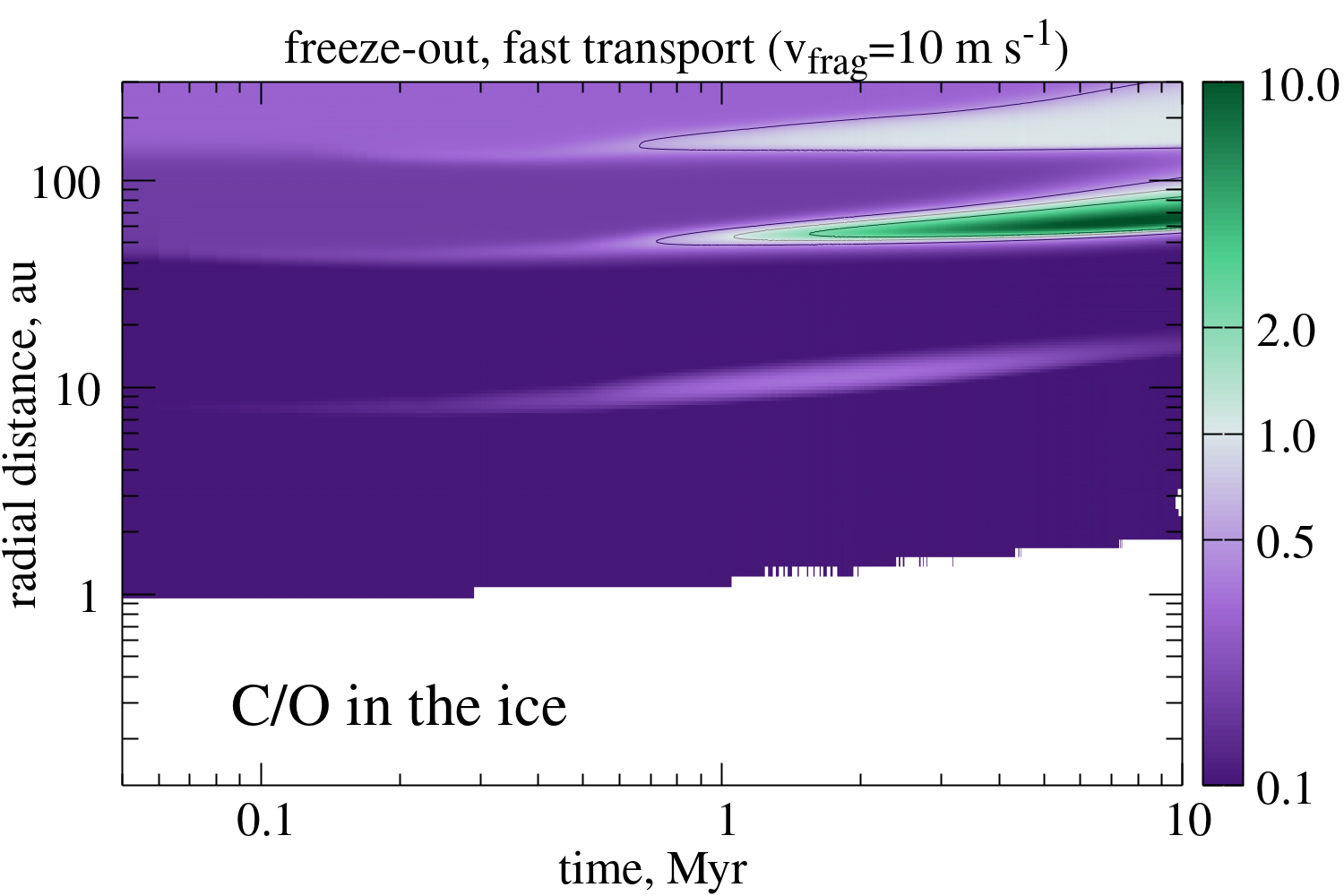}
    \caption{Evolution of the radial distributions of C/O ratios in the gas (left) and in the ice (right). Models with no transport, slow dust transport ($v_{\rm frag}=1$\,m~s$^{-1}$) and fast dust transport ($v_{\rm frag}=10$\,m~s$^{-1}$) are shown, with baseline or freeze-out only chemistry. The regions where the total abundance of~C and~O in gas or ice is lower that 0.001 of their initial total abundance are shown in white. The contours mark C/O values of 0.5, 1 and 2.}
    \label{fig:c2o_evolution_baseline_no_gap}
\end{figure*}

To characterise the elemental composition of gas and ice, we use the C/O ratios as a diagnostic. Most abundant molecules observed in protoplanetary discs contain C, O or both, and the value of the C/O ratio allows to distinguish between carbon- and oxygen-rich chemistry regimes. The C/O ratio in the gas (ice) phase is defined as the total number of carbon atoms present in species in the gas (ice), divided by the total number of oxygen atoms in these species. We calculate the C/O ratios in the gas and in the ice at every radial distance and show the evolution of their radial distributions in Figure~\ref{fig:c2o_evolution_baseline_no_gap}.

In the model with no transport, the amount of~C and~O at a given radial distance is fixed, so the changes in the C/O ratio can only arise from conversion between ice and gas phase. Initially, the C/O ratio has a step-like structure determined by snowlines \citep[cf.][]{2011ApJ...743L..16O}, and it evolves as the species are converted to each other. The main features of this model are oxygen-rich gas in the inner disc (inside the water snowline) and carbon-rich gas and oxygen-rich ice in the outer disc. The gas-phase C/O ratio outside 10\,au (the \ce{CO2} snowline) is initially above~1, the gas composition being dominated by \ce{CO} and \ce{CH4}. It decreases to~$1$ after $\approx1$\,Myr, when methane depletion removes carbon from the gas phase. Consequently, the ice-phase C/O in the outer disc gradually increases from $\sim0.1$ to the initial volatile ratio of $0.29$. By the age of $\approx8$\,Myr, almost all carbon and oxygen beyond 20\,au is removed from the gas by chemical reactions.

The addition of dust and gas transport significantly affects the C/O ratio at all radial distances already at the age of $\sim0.1$\,Myr. Water ice brought inside 1\,au by pebble drift decreases the gas-phase C/O ratio from 0.29 to $0.21$ in the inner disc due to increasing oxygen content there. Between the \ce{H2O} and \ce{CO2} snowlines, the C/O in the gas starts as~1, due to equal initial abundances of \ce{CO2} and \ce{CH4}, then it decreases and approaches 0.5 due to \ce{CO2} enrichment. Similarly, gas-phase C/O increases in a region just inside the \ce{CH4} snowline. Accumulation of volatiles at the snowlines leads to additional peaks in the C/O ratios both in the gas and in the ice \citep[see, e.g.,][]{2017MNRAS.469.3994B}. Unlike in the model with no transport, here the gas does not become depleted of C and O inside $\approx200$\,au even at later stages, thanks to the continued influx of volatiles from the outer disc.

The choice of $v_{\rm frag}$ determines the timescale at which the resulting C/O ratio distribution evolves. The slow dust transport model is qualitatively very similar to the case without transport, until the age of a few Myr. The most significant differences at these times are achieved for the gas-phase C/O and do not exceed $\sim30$\%. In the fast transport model, the variations in the C/O compared to the model with no transport are stronger. The evolution is only similar before 1\,Myr; after that, all the gas in the disc becomes more carbon rich, including in the inner 1\,au, and the C/O ratio in the ice also approaches~1 in most of the disc. This is a result of pebble drift efficiently bringing in more carbon-rich ices, including \ce{CO2} and hydrocarbons formed from the products of CO and methane destruction. In the absence of dust traps, efficient radial drift depletes the disc of most of its oxygen on timescales of $\approx3$\,Myr. 

Ultimately, it is the combination of chemical reactions and radial drift that determines the C/O ratio distribution. In the freeze-out only model (lower panels of Figure~\ref{fig:c2o_evolution_baseline_no_gap}), the gas becomes much more carbon rich at a few Myr, while the ice is still O-rich, except at the snowlines of C-bearing species. Chemistry creates more complex carbon-bearing species (e.g., \ce{CH2CHCCH}, \ce{C3O}, \ce{C2H6}), particularly in the ice, enriching the ice phase with carbon, increasing its C/O ratio. It also facilitates carbon depletion of the disc as a whole, resulting in a lower C/O ratio of the gas ($\approx1$ vs~$>2$ in the freeze-out only model).

\subsection{Gas-phase C/O ratio in the inner disc}
\label{sec:baseline_chamistry}

Observations with JWST provide column densities of multiple molecules in the inner disc, but we cannot directly compare them with our modelling results. Emission of abundant molecules, particularly water, is typically optically thick and traces the composition of the upper layers \citep{2018A&A...618A..57W,2023A&A...679A.117G}. The composition can be relatively reliably characterised by column density ratios \citep[e.g.,][]{2015A&A...582A..88W,2024ApJ...977..173C,2026ApJ..1000..217D}. Our modelling simulates the midplane composition, which is not directly traced by observations. However, it is likely that the elemental abundances are vertically homogeneous, because the turbulent mixing allows efficient exchange of matter between the midplane and upper layers, \citep[see, e.g.,][]{2024ApJ...977..173C,2025MNRAS.537..691H}. This can be seen from a comparison between the mixing timescale $t_{\rm mix}$ and accretion timescale $t_{\rm acc}$. At a given distance $R$ they can be assessed from the local angular velocity $\Omega$, radial velocity $v_R$ and gas scale height $H$ as
\begin{equation}
    t_{\rm mix} = \frac{1}{\alpha \Omega} \hspace{1mm} < \hspace{1mm}
t_{\rm acc} = \frac{R}{v_R} \sim \frac{R^2}{\alpha c_{\rm s} H} = \frac{R^2}{\alpha \Omega_{\rm K} H^2} = \left( \frac{R}{H} \right)^2 t_{\rm mix}
\end{equation}
Here we used the fact that $c_{\rm s} = H \Omega_{\rm K}$, and the local mass accretion rate $\dot{M}$ can be expressed either through gas velocity or $\alpha$-viscosity $\dot{M} = 3\pi \alpha c_{\rm s} H \Sigma_{\rm gas} = 2\pi R v_R \Sigma_{\rm gas}$, where $\Sigma_{\rm gas}$ is gas surface density. As a typical scale height to radial distance ratio is of the order of 0.1, mixing should proceed much faster than radial transport. Therefore the comparisons between bulk elemental ratios are likely more meaningful than those of molecular abundances derived from our models, because the molecular composition of the upper layers is additionally affected by photo-reactions~\citep{2021ApJ...909...55A}. Models including both radial and vertical transport, as well as chemical evolution, would be useful to properly assess the relation between the observed upper layers and the bulk chemical composition. Despite the uncertainties in chemical models and emission of different molecules coming from different vertical locations \citep{2018A&A...618A..57W}, gas-phase C/O ratios in the inner disc can be inferred from flux ratios of particular molecules \citep{2021ApJ...909...55A}, e.g., the HCN/\ce{H2O} ratio has been shown to be correlated with the C/O ratio \citep{2011ApJ...743..147N,2013ApJ...766..134N}. Thermochemical modelling shows that observations of \ce{C2H2} and larger hydrocarbons with JWST is evidence for bulk $\mathrm{C/O}>1$ \citep{2024A&A...689A.231K,2025A&A...699A.194A}.

We adopt 1\,au as a dividing distance between the inner and the outer disc. Analysis of JWST observations of molecules in the inner disc also suggests that emission is coming from a region of approximately 1\,au size \citep{2023ApJ...957L..22B,2023A&A...679A.117G,2024ApJ...963..158P,2024ApJ...964...36R,2025AJ....169..184S,2025A&A...701A.239S}. This is also the approximate location of the water snowline in our model. Inside this distance, most molecules are in the gas, except for the most refractory organic species. The initial C/O ratio for the adopted composition is 0.55 in total, and 0.29 for the volatiles, i.e. excluding the dust-grain cores (silicates and refractory carbon). In this work, we only consider the volatile species, without taking into account the composition of dust-grain cores that sublimate at significantly higher temperature. The gas-phase C/O in the inner disc is close to the total volatile C/O ratio, except for the case of active carbon grain destruction, which is considered in Section~\ref{sec:carbon_destruction}.

\begin{figure*}
\includegraphics[width=\columnwidth]{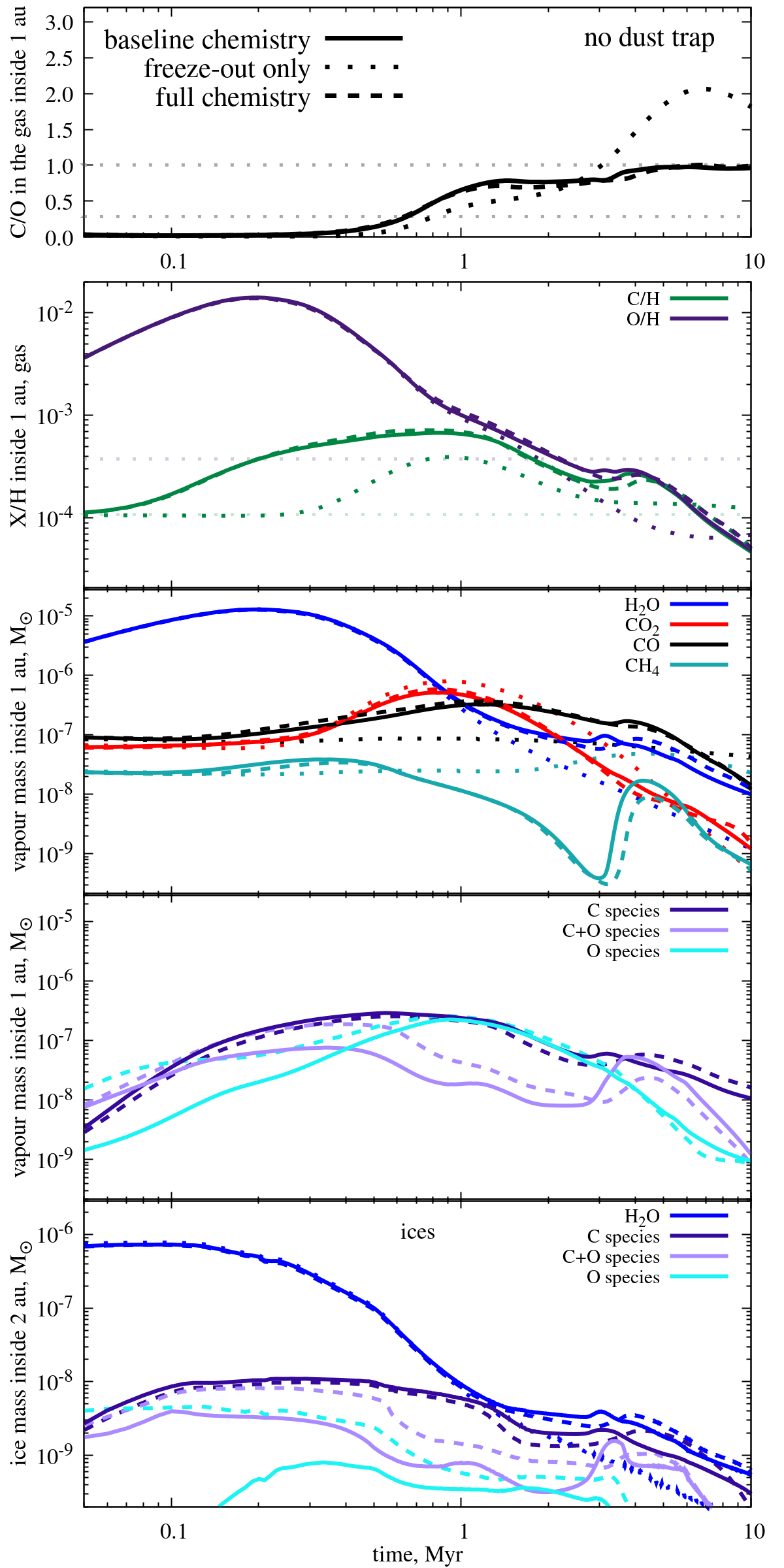}
\includegraphics[width=\columnwidth]{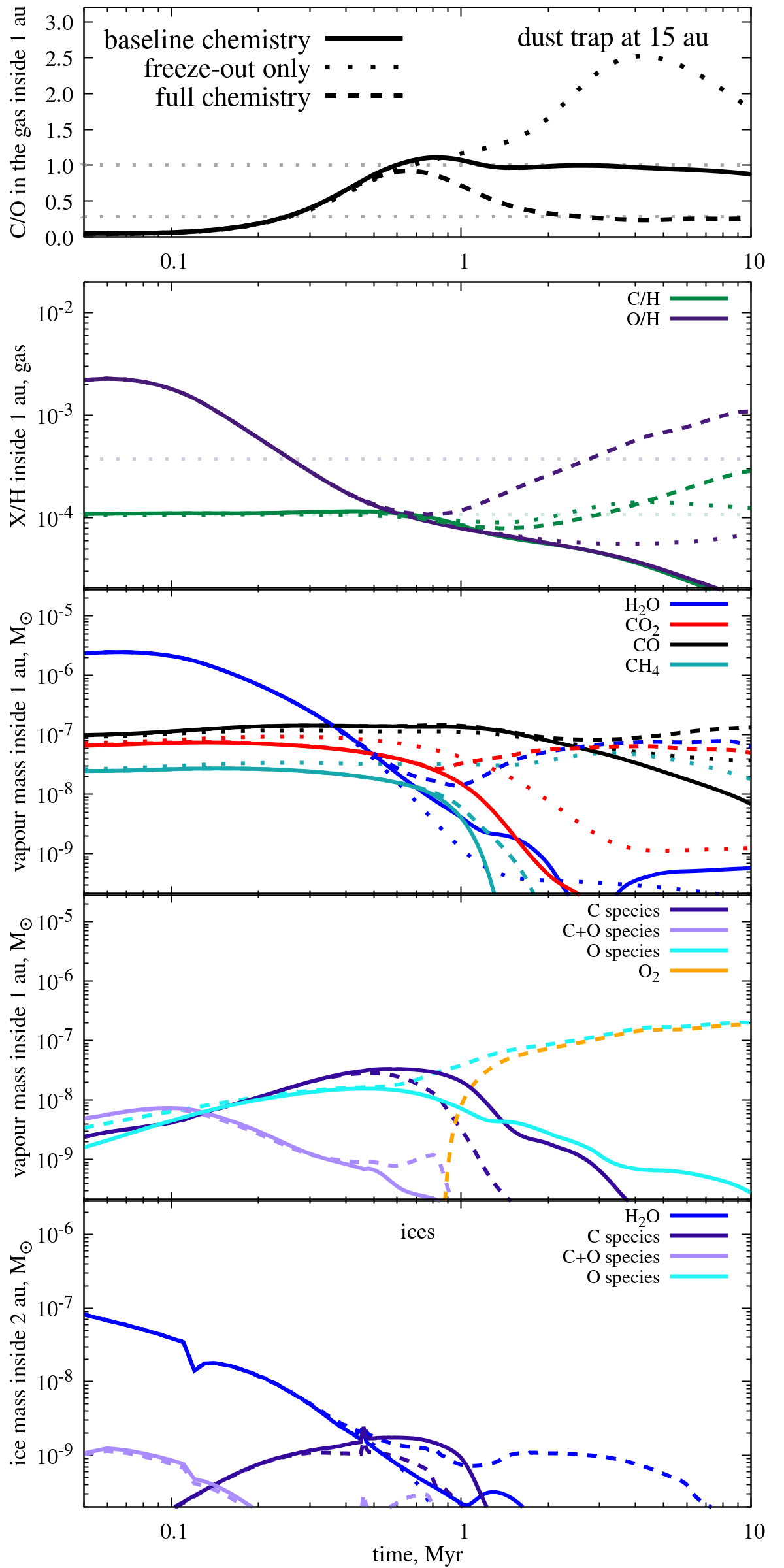}
    \caption{Evolution of integrated masses of main gas-phase volatiles and C/O ratios in the inner disc in the models with fast transport ($v_{\rm frag} = 10$\,m~s$^{-1}$). Models with baseline chemistry, freeze-out only, and full chemistry are marked by solid, dotted, and dashed lines, respectively. Left: smooth disc (no dust trap), Right: planet-induced dust trap at 15\,au. Purple, lavender, and cyan lines show cumulative amount of groups of species: C~species, C+O~species, and O~species, respectively. \ce{O2} is included in O~species. Transparent dotted lines mark the reference levels: 1~and 0.29~(initial value) for the C/O ratio, and solar values for the C/H (green) and O/H (purple) ratios.} 
    \label{fig:volatiles_1au_c2o}
\end{figure*}

Figure~\ref{fig:volatiles_1au_c2o} shows how the chemical composition of the inner disc evolves in models with different chemistry regimes, with or without dust traps. In the upper panels, we show the integrated gas-phase C/O ratio in the inner 1\,au as the observationally relevant parameter. Then we plot the metallicity of this gas, i.e. the C/H and O/H elemental ratios that shape the C/O ratio in this region.
To demonstrate the contribution of different gas-phase volatiles, we show the masses of the initially present species \ce{H2O}, \ce{CO2}, CO and \ce{CH4}, and the newly formed carbon- and oxygen-bearing species in separate panels. ``C~species'' include various hydrocarbons that contain no oxygen, such as \ce{C_nH_m}, \ce{H2CS}, etc. ``C+O~species'' mostly comprise O-bearing COMs, such as \ce{CH3CHO}, \ce{CH3OH}, \ce{HCOOCH3}, \ce{CH2OHCHO}, as well as some simpler species, such as \ce{H2CO}. ``O~species'' refer to oxygen-bearing molecules containing no carbon, including SO, \ce{SO2}, \ce{O2}, as well as NO, \ce{NH2OH} and other N-bearing species. The initially present molecules (CO, \ce{CH4}, \ce{CO2} and \ce{H2O}) are excluded from these groups. Finally, in the lower panels of Figure~\ref{fig:volatiles_1au_c2o}, we demonstrate the masses of the ice-phase species in the inner 2\,au. They are not included in the inner disc C/O ratio, but they show the composition of the volatiles arriving to the inner 1\,au in the form of ice on drifting dust grains. We analyse these results in the following sections for models with and without (i.e. smooth disc) dust traps.

\subsubsection{Smooth discs}
\label{sec:c2o_in_no_trap_discs}

In the freeze-out models, \ce{H2O}, \ce{CO2}, CO and \ce{CH4} are the only carbon- or oxygen-bearing volatiles. Within the initial $\sim100$\,kyr, the C/O in the inner disc decreases, as water arrives to the inner disc with drifting pebbles. More volatile \ce{CO2} has the snowline farther from the star, at $\approx10$\,au, and the excess \ce{CO2} arrives later, when the water reservoir already starts to be exhausted, leading to an increase in the C/O ratio. After that, carbon-rich material and even more volatile CO and \ce{CH4} arrive in the inner disc, causing the C/O to increase above unity at $\gtrsim3$\,Myr. This sequence of volatile enrichment by pebble drift is similar to what was reported in the previous studies \citep{2017MNRAS.469.3994B,2021ApJ...921...84K,2023ApJ...954...66K,2023A&A...677L...7M,2024A&A...686L..17M}. 

During the first $\approx2$\,Myr, the inclusion of chemistry leads to a slight increase in the inner disc C/O ratio in the model without dust traps (left panels of Figure~\ref{fig:volatiles_1au_c2o}). With chemistry, less volatile C- and O-carriers are formed all over the disc, and they reside predominantly in the ice phase, as described in Section~\ref{sec:baseline_chamistry}. These new species continue to chemically evolve as they are being delivered with dust grains to the inner disc, where they are sublimated to the gas, and chemically reprocessed mainly to CO. This increases the C/H ratio at early times ($\lesssim0.5$\,Myr), but the C/O remains close to zero, as the composition is dominated by the  pre-existing species such as water, CO and \ce{CO2}.
This is consistent with earlier studies concluding that at moderate turbulence and ionisation rate, dust drift has a more significant effect on chemical composition than chemical reactions which have a longer timescale \citep{2019MNRAS.487.3998B,2018A&A...613A..14E}. 

At $\approx1$\,Myr, water ceases to be the dominant volatile in the disc. After dust drift depletes the disc of \ce{H2O}, ices become carbon-rich (see the third row of panels in Figure~\ref{fig:c2o_evolution_baseline_no_gap}). This carbon enrichment of models with chemistry relative to freeze-out only models is caused by the cosmic ray-induced destruction of methane and to a lesser extent, CO, in the outer disc regions, followed by the formation of more complex molecules. The species formed at $10-100$\,au that inherit the carbon from \ce{CH4} and CO include a variety of C~species and C+O~species (see Section~\ref{sec:no_transport}). An important source of oxygen in the inner disc is presented by O~species draining oxygen from water and \ce{CO2}. They are dominated by SO\footnote{Unlike in the cold interstellar environment, SO in protoplanetary discs is not necessarily associated with shocks. At temperatures above $\approx50$\,K, it stays in the gas phase and in combination with CS can be used as a tracer of C/O ratio \citep{2021ApJS..257...12L}.} and other S-bearing molecules forming from \ce{H2S}, which is initially the only S-bearing species with an initial abundance of $1.32 \times 10^{-5}$ relative to H~nuclei. This initial abundance implies that all the available sulphur is in the volatile reactive form. However, there is observational evidence that most of sulphur in protoplanetary discs is stored in some refractory form \citep{2019ApJ...885..114K}, and other disc models assume lower sulphur abundances to account for this depletion \citep[e.g.,][]{2002A&A...386..622A,2011A&A...535A.104D,2019ApJ...876...72L}. Upon arrival to the inner disc, more species are formed, such as \ce{H2CS}, \ce{C9H2}, \ce{NH2CHO}, \ce{CH3CCH}, \ce{SO2}.

After $\approx3$\,Myr, the inner disc composition changes qualitatively. In the freeze-out only model the C/O ratio continues to grow above~1 as methane starts to arrive in the inner disc, enriching it with carbon. At the same time, in the baseline chemistry models, the C/O ratio is still around 0.7, slowly approaching~1 by 6\,Myr. In the models with chemistry, the methane delivery does not happen, as the methane no longer exists; instead, the delivery of carbon and oxygen to the inner disc is dominated by gas-phase CO, with minor contributions from methanol and other COMs.

The transformation of molecules within 1\,au does not affect the C/O ratio, although it affects the molecular composition. Particularly, at $T>700$\,K (inside 0.3\,au) collisional dissociation and two-body reactions in the gas phase destroy \ce{CO2} and HCN, as well as newly formed hydrocarbons and COMs. Products of these reactions form other molecules, mostly CO, \ce{CH4}, \ce{H2O}, and a variety of C~species which are the most stable at these densities and temperatures. Destruction of refractory organics at high temperatures could release additional carbon from the dust-grain cores and contribute to the C/O ratio in the gas in this inner region~\citep[e.g.,][]{2024MNRAS.535..171P,2025A&A...699A.227H}. We consider this effect in Section~\ref{sec:carbon_destruction}.

In a smooth disc without dust traps, the inner disc composition in the baseline chemistry models is very similar to the full chemistry models (which include cosmic ray photo-processing of ices). As shown in Figure~\ref{fig:no_transport_radial}, one of the main effects of full chemistry is the dissociation of water that releases some oxygen into other O-bearing species. In the full chemistry model, more O-rich species with binding energies lower than water are formed, such as SO and \ce{SO2}. This leads to more oxygen retained in the gas phase, and thus less oxygen lost due to rapid dust transport. Overall, it results in a slightly more oxygen-rich disc and marginally lower gas-phase C/O ratios in the inner disc.

\subsubsection{Discs with a dust trap}
\label{sec:c2o_in_discs_with_traps}

The inclusion of a dust trap prevents pebbles from reaching the inner disc, reducing the delivery of ices. The evolution of the gas-phase C/O ratio and the main volatile components in the models with a dust trap at 15\,au is shown in the right panels of Figure~\ref{fig:volatiles_1au_c2o}. Among the initially present species, the dust trap interrupts the delivery of water and \ce{CO2}, but does not affect the more volatile CO and \ce{CH4}. This leads to earlier growth of the C/O ratio compared to the smooth disc models. A dust trap situated inside the CO and \ce{CH4} snowlines but outside the \ce{H2O} and \ce{CO2} snowlines suppresses oxygen transport by pebbles, favouring carbon-rich inner discs, which leads to higher maximum values of the C/O achieved earlier in all chemical regimes, compared to the case with no trap. Evolution of radial C/O ratio distributions in models with dust traps is shown in Appendix~\ref{sec:c2o_with_trap}.

In models with chemical reactions, the transport of complex organics formed from the products of methane destruction is also interrupted by the dust trap, and most of the carbon from beyond 15\,au never reaches the inner disc. After $\approx1$\,Myr, when dust radial drift depletes the region interior to the trap of most of its solid-phase volatile content, carbon enrichment of the inner disc stops, and the C/O ratio cannot achieve values of $>1$. CO becomes the dominant volatile among C- and O-bearing species in both the freeze-out only and baseline chemistry models.

In the full chemistry model, cosmic-ray induced photo-processing of ices in the dust trap leads to a qualitatively different picture, which is discussed in Section~\ref{sec:CRPHOT_ices_trap}. The C/O ratio in the inner disc sharply decreases to $\approx0.25$ at $>2$\,Myr. The decrease is caused by additional enrichment of the inner disc with molecular oxygen formed in the dust trap in a reaction sequence triggered by cosmic ray induced photo-reactions in the ice phase (see Appendix~\ref{sec:O2_formation} for details). As a result, after 2\,Myr, \ce{O2} and CO become the most abundant volatiles in the inner 1\,au in this model. In the very inner region (inside 0.3\,au) these molecules are thermally reprocessed in the gas-phase, forming water and \ce{CO2}.

\subsubsection{Full chemistry in a dust trap}
\label{sec:CRPHOT_ices_trap}

In the full chemistry model, water ice in a dust trap at 15\,au can be dissociated to form the hydroxyl radical OH, which reacts with other surface species, eventually producing \ce{O2} ice \citep[see Appendix~\ref{sec:O2_formation} and][]{2018A&A...613A..14E}. The binding energy of \ce{O2} (892\,K, comparable to CO, 855\,K) is too low for it to stay on dust surface at 15\,au ($\approx53$\,K), so it immediately escapes to the gas phase, where it is instead being advected with the gas and can reach inner disc regions. This mechanism provides a way to release the trapped oxygen and deliver it to the inner disc. The same \ce{O2} formation is also happening in the smooth disc with no traps in the full chemistry model, but it is too slow to make a big difference when transport is accounted for. We do not see any additional \ce{O2} in the inner disc (left panels of Figure~\ref{fig:volatiles_1au_c2o}). With freely drifting pebbles, these reactions are much less efficient, as this water ice reservoir is spread throughout the disc and quickly depleted by radial drift.

The potential for efficient oxygen production in the dust trap makes our predictions for the C/O ratio in the inner disc particularly sensitive to the efficiency of cosmic ray induced photo-reactions in the ice phase. Due to the limitations of the two-phase approach to surface chemistry in the model, the full chemistry model is likely overestimating their contribution, while the baseline chemistry model may underestimate it (see discussion in Section~\ref{sec:O2_observation_discussion} and Appendix~\ref{sec:O2_formation}). We tested models with different efficiency factors for these reactions, by multiplying their rates by a $C_{\rm iCR}$ factor. $C_{\rm iCR}=1$ corresponds to full chemistry, and  $C_{\rm iCR}=0$ corresponds to baseline chemistry models. We test intermediate values of  $C_{\rm iCR}=10^{-3}$, $10^{-2}$ and $10^{-1}$ that correspond to different level of involvement that the species produced in the bulk mantle can have in surface chemistry. Given that the typical thickness of ice mantles is around $10^2-10^3$ monolayers assuming compact spherical grains, $C_{\rm iCR}=10^{-3}-10^{-2}$ would imply that the reaction products from the bulk of ice mantle have no access to surface chemistry, while at $C_{\rm iCR}=10^{-2}-10^{-1}$ they would partially diffuse to the surface or be close enough to each other to react in the bulk. The $C_{\rm iCR}$ coefficient can also be interpreted as a measure of inhomogeneity of dust grain surface, with $C_{\rm iCR}=1$ corresponding to a fractal and porous ice mantle structure.

\begin{figure}
\includegraphics[width=\columnwidth]{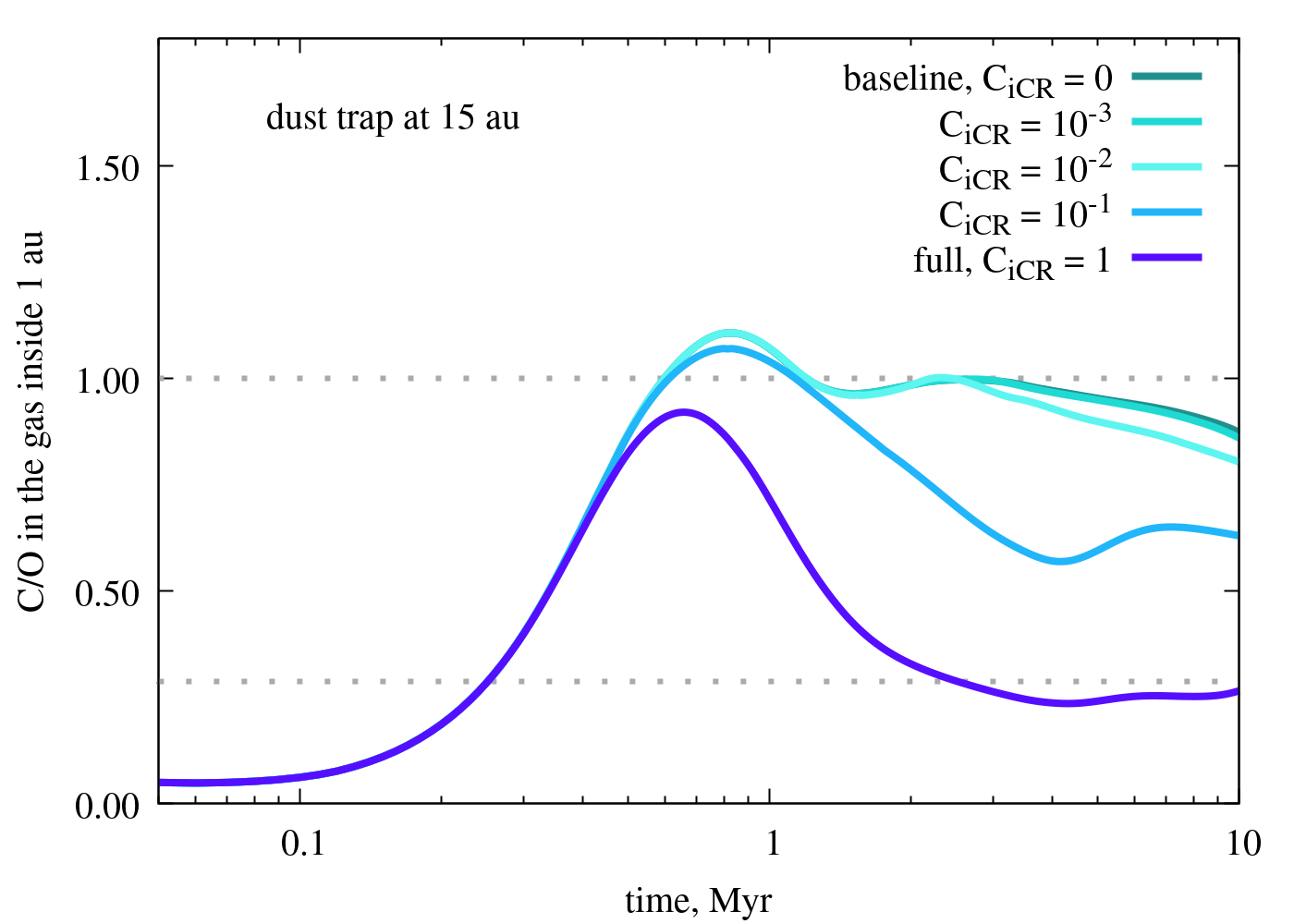}
    \caption{C/O ratio in the inner disc regions in the full chemistry models with different efficiencies of cosmic ray induced dissociation of ices.}
    \label{fig:CRPHOT_mantle_coefficient}
\end{figure}

The simulated C/O ratios in the inner disc at different values of $C_{\rm iCR}$ are shown in Figure~\ref{fig:CRPHOT_mantle_coefficient}. At reduction factors of $C_{\rm iCR}=10^{-3}$ and $10^{-2}$, the C/O ratio in the inner disc is marginally lower than in the baseline chemistry model, and significant oxygen enrichment is only seen in the $C_{\rm iCR}=10^{-1}$ case. The amount of oxygen that reaches the inner disc at 5\,Myr is roughly proportional to $C_{\rm iCR}$, leading to the corresponding change in the C/O ratio. For significant \ce{O2} production and qualitatively lower C/O ratios, about 10\% of the dissociation products in the ice mantle should be able to take part in surface chemistry.

\subsection{Efficiency of radial drift and the location of the dust trap}
\label{sec:trap_location}

\begin{figure}
\includegraphics[width=1\columnwidth]{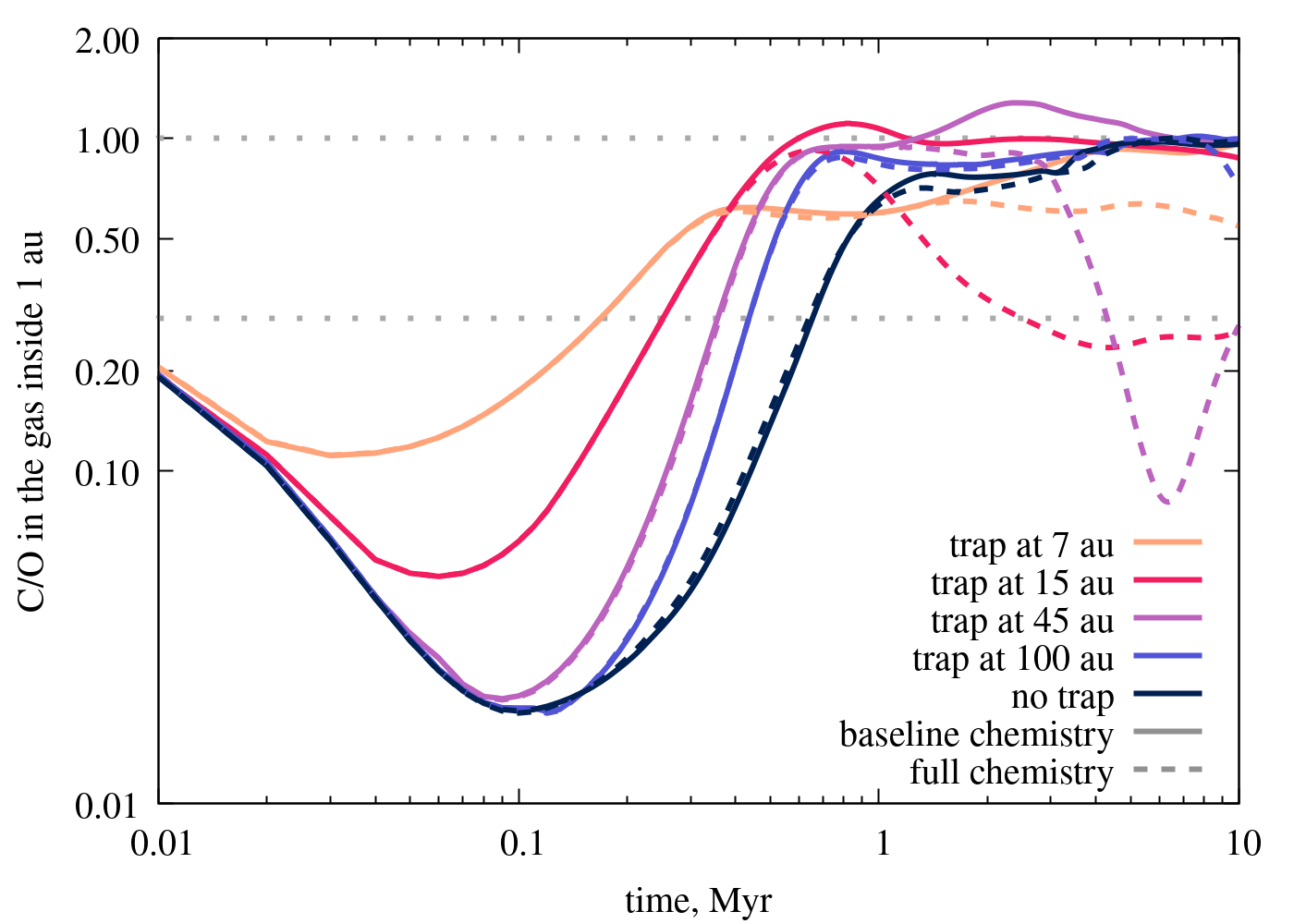}
    \caption{Gas-phase C/O ratios in the inner disc in models with baseline and full chemistry and different locations of the dust trap. }
    \label{fig:c2o_trap_location}
\end{figure}

To test the effect of the dust trap's location, we put the planet at 5, 30, and 70\,au from the star, resulting in a dust trap at 7, 45, and 100\,au, respectively, versus the trap at 15\,au in the main gap models. The inner disc C/O ratios for different trap locations are shown in Figure~\ref{fig:c2o_trap_location}.

The location of the dust trap has the most significant effect on the C/O ratios before $\approx1$\,Myr when pebble flux is the highest, as it affects the timing and the degree of oxygen delivery. As was previously demonstrated by \citet{2021ApJ...921...84K} and \citet{2024ApJ...977...21E}, a trap located further away from the star results in longer and stronger water enrichment of the inner disc. We note that modelling with more sophisticated models \citep[e.g. \texttt{DustPy};][]{2022ApJ...935...35S}, shows lower leakiness in \texttt{two-pop-py}, which could weaken these effects \citep{2026arXiv260411925H}. After 1\,Myr, the baseline chemistry models are more similar, with the most noticeable differences in the models with the trap at~7 and 100\,au. The dust trap at 7\,au is within the \ce{CO2} snowline, so it does not interrupt the influx of \ce{CO2} to the inner disc, making the C/O ratio closer to 0.5. The trap at 100\,au is beyond both the CO and \ce{CH4} snowlines, so it blocks the delivery of CO and \ce{CH4} ices via pebble drift. However, only about a third of the disc material is beyond 100\,au, which makes this case qualitatively similar to the model with no traps.

With full chemistry, formation of \ce{O2} in the dust trap and the subsequent oxygen enrichment of the inner disc are also affected by the trap location: it is only efficient in models with a dust trap at intermediate distances (15 and 45\,au). A close in dust trap (7\,au) produces a substantial amount of molecular oxygen after 7\,Myr, slightly decreasing the C/O ratio of the inner disc to $\approx0.6$. With the trap at 100\,au, the inner disc enrichment does not happen within 10\,Myr, with C/O ratio staying around unity at later times. However, some amount of \ce{O2} is still produced in all full chemistry models with dust traps. Potential observational constraints on the efficiency of dissociation of ices by cosmic ray induced photons is discussed in Section~\ref{sec:O2_observation_discussion}.

\subsection{Ionisation rate}
\label{sec:ionisation_rate}

\begin{figure*}
\includegraphics[width=\columnwidth]{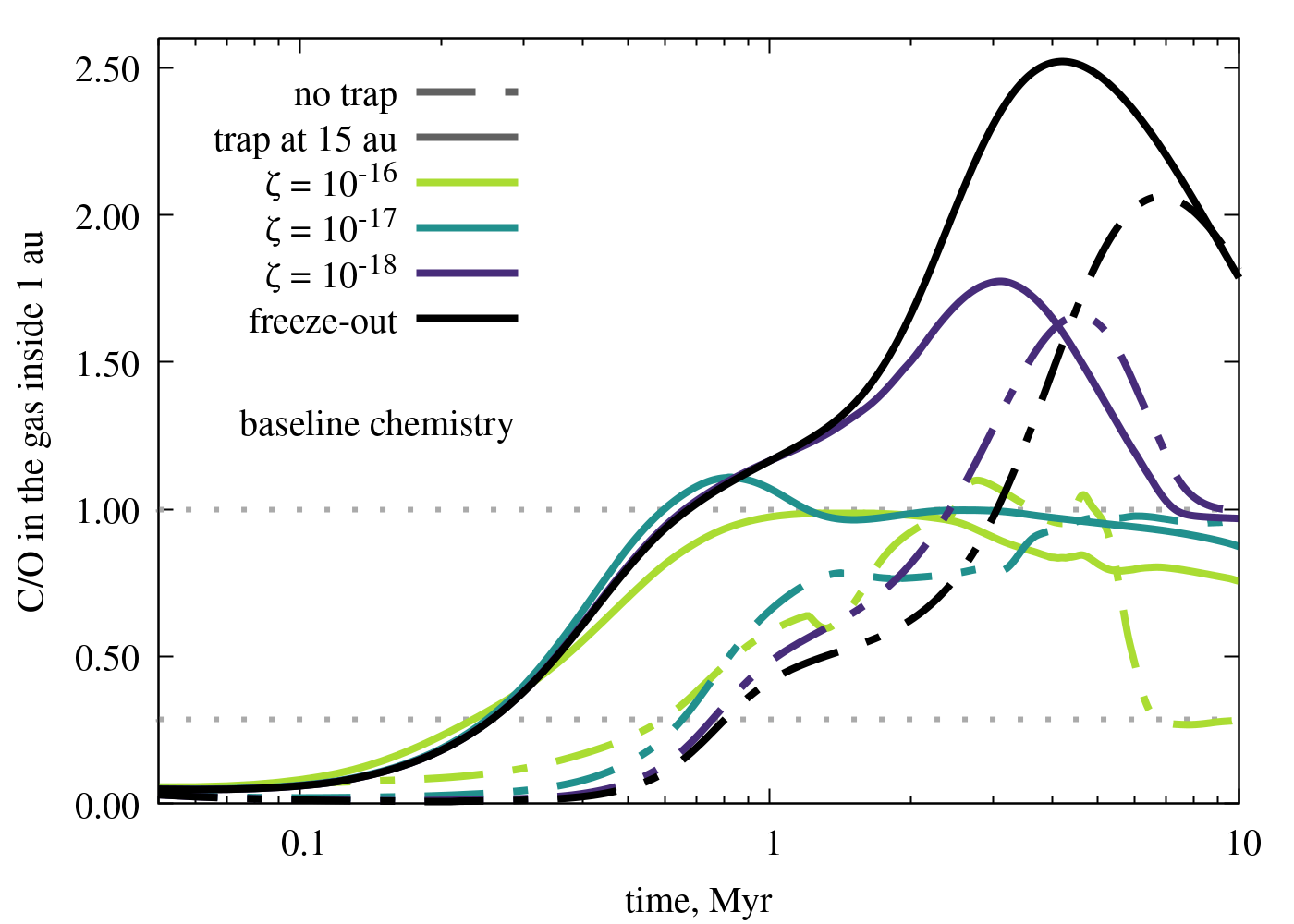}
\includegraphics[width=\columnwidth]{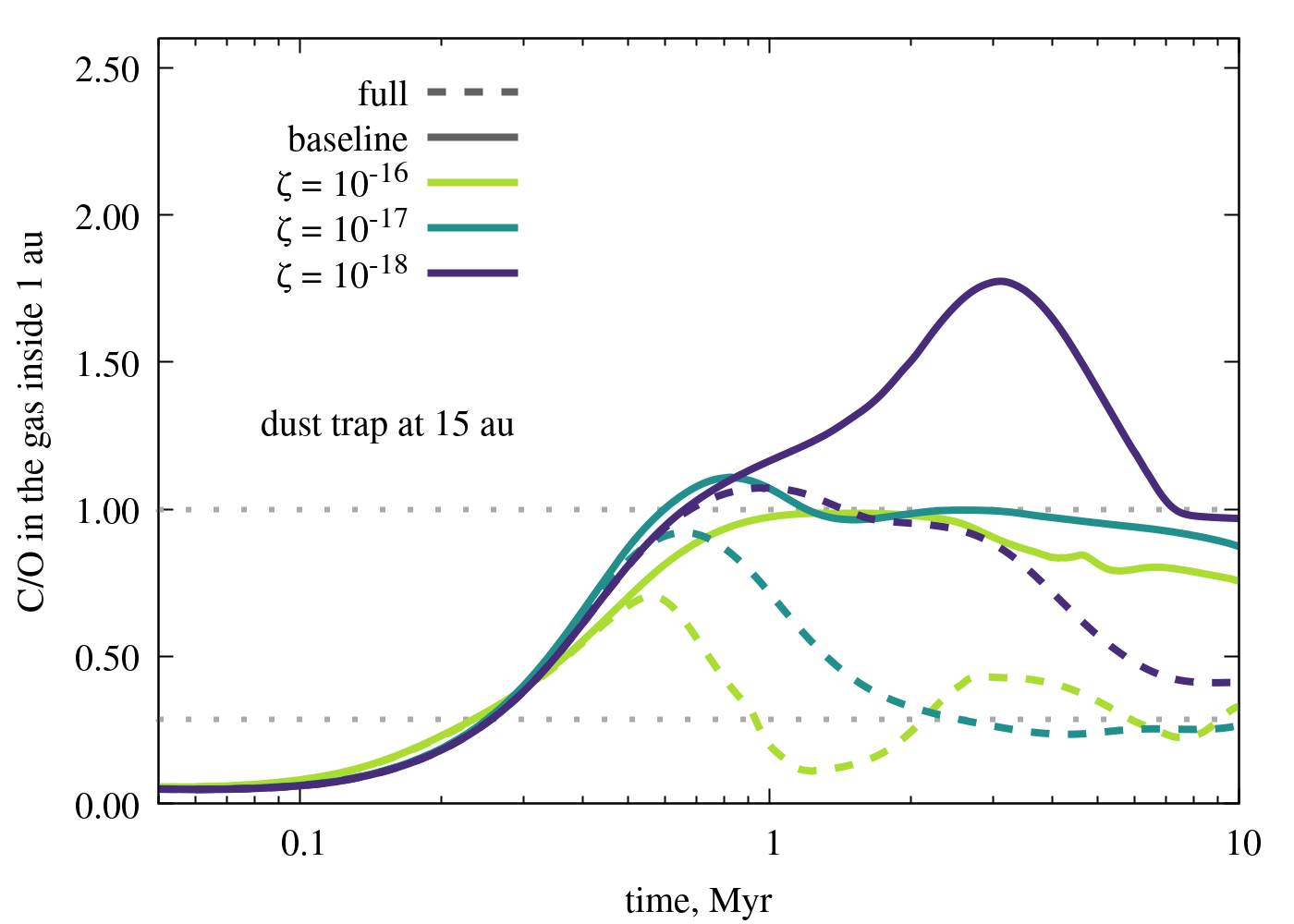}
    \caption{C/O ratio in the inner disc regions in models with different cosmic ray ionisation rates. Left panel shows baseline chemistry models compared with the freeze-out model (black), with (solid lines) and without (dot dashed lines) dust traps. Right panel shows the baseline (solid lines) and full (dashed lines) chemistry models with a dust trap at 15\,au.}
    \label{fig:CR_basechem_fullchem}
\end{figure*}

Cosmic ray ionisation is the main driver of chemical evolution in the disc midplane. The incident cosmic ray ionisation rate is not very well known for protoplanetary discs, and it may vary between star forming regions and due to the local circumstellar environment. The available constraints report values both lower \citep[$10^{-19}$\,s$^{-1}$][]{2015ApJ...799..204C} and higher \citep[up to $0.5-1 \times 10^{-16}$\,s$^{-1}$,][]{2006PNAS..10312269D,2021ApJS..257...13A} than the universally assumed  value for local interstellar clouds of $1.3 \times 10^{-17}$\,s$^{-1}$. The model of cosmic ray propagation in circumstellar discs by \citet{2018A&A...614A.111P} provides even higher cosmic ray ionisation rates.

The left panel of Figure~\ref{fig:CR_basechem_fullchem} demonstrates the effect of different CR ionisation rates in models with baseline chemistry. At lower $\zeta$, the evolution is similar to the freeze-out case both with and without the trap. The dissociation of methane is less efficient, leading to later enrichment of the inner disc with carbon-bearing species. At low cosmic ray ionisation, the effect of chemistry is minor, in agreement with previous findings \citep{2019MNRAS.487.3998B}. However, the maximum reached C/O value is sensitive to the initial abundance of methane, which is relatively high in the initial composition we adopt here \citep[see][]{2023A&A...677L...7M,2025A&A...699A.227H}. 

At high $\zeta= 10^{-16}$\,s$^{-1}$, the destruction of methane is more efficient and starts affecting C/O ratios earlier, at a few kyr timescale. By 1\,Myr, the carbon-rich material is already mostly lost to accretion, while \ce{CO2} is still abundant, which results in a C/O ratio below~1. After 6\,Myr, the C/O ratio sharply decreases due to destruction of CO in the gas-phase, leaving water and the rest of COMs as main carriers of carbon and oxygen. This destruction also happens in the fiducial model, but less efficiently before 10\,Myr.

This result is qualitatively different to the conclusions of \citet{2025A&A...701A.239S}, who found that a high ionisation rate leads to more carbon enrichment of the inner disc. This different conclusion comes from the differences in the chemical networks. The main process responsible for carbon enrichment in the models of \citet{2025A&A...701A.239S} is the conversion of CO to \ce{CH4} through reactions of the \ce{C+} released from CO with molecular hydrogen. In our model, in addition to methane production, \ce{C+} destroys \ce{CH4}, typically at a faster rate than it is produced, until the methane abundance becomes very low ($\sim10^{-7}$). This complements the \ce{CH4} destruction pathway from direct CRPHOT dissociation, which \citet{2025A&A...701A.239S} consider in a subset of their models. As a result, in our model, methane production from \ce{C+} is a factor of a few less efficient than its destruction. Since both formation and destruction of \ce{CH4} are triggered by the same \ce{C+} ions, we see faster methane destruction at high ionisation rates. Overall, our model predicts net destruction of \ce{CH4} via \ce{C+}, while the model of \citet{2025A&A...701A.239S} predicts net production. Thus, a high ionisation rate leads to more efficient \ce{CH4} production in their model, creating the opposite trend with $\zeta$. The chemistry of methane as the main primordial carbon-dominated volatile plays a crucial role in the inner and outer disc composition and is very sensitive to the ionisation rate. The global results of chemical modelling are determined by the main production and destruction routes included in the reaction network.

For the models with full chemistry, we also find that the C/O ratios in the inner disc are lower at higher $\zeta$ (see right panel of Figure~\ref{fig:CR_basechem_fullchem}). The formation of \ce{O2} also depends on the CR~ionisation rate, which leads to more oxygen enrichment and an earlier decrease in the C/O ratio at higher ionisation rate. As in the baseline chemistry models, a high $\zeta$ also leads to more methane destruction and consequently weak late carbon enrichment, as well as additional CO destruction after 6\,Myr.

\subsection{Carbon grain destruction}
\label{sec:carbon_destruction}

Gas in the inner disc regions can be additionally enriched in carbon if the refractory carbon from solid dust grains is chemically processed. Normally, amorphous carbon would sublimate at temperatures above 1000\,K, but dust grains contain multiple chemically diverse carbonaceous components, that could be transferred to the gas at different temperatures \citep{2017A&A...606A..16G}. Pathways include pyrolysis (i.e. thermal decomposition in the absence of oxygen) of refractory organic material at $300-500$\,K \citep{2003ApJ...592.1252N,2017A&A...606A..16G}, destruction by~O and~OH atoms at high temperatures and UV radiation field \citep{2002A&A...390..253G,2010ApJ...710L..21L,2019ApJ...870..129W}. Based on the data from Solar System comets, a significant fraction of carbon in the ISM should be present in the dust component \citep{1994GeCoA..58.4503F,2017A&A...606A..16G}. This amount of additional carbon could contribute to high C/O ratios in the inner disc regions \citep{2021SciA....7.3632L,2024MNRAS.535..171P,2025A&A...699A.227H}. An important aspect of this process its irreversibility: once released, the carbon from grains is transformed into a more volatile species that would condense only at much lower temperature.

\begin{figure}
\includegraphics[width=1\columnwidth]{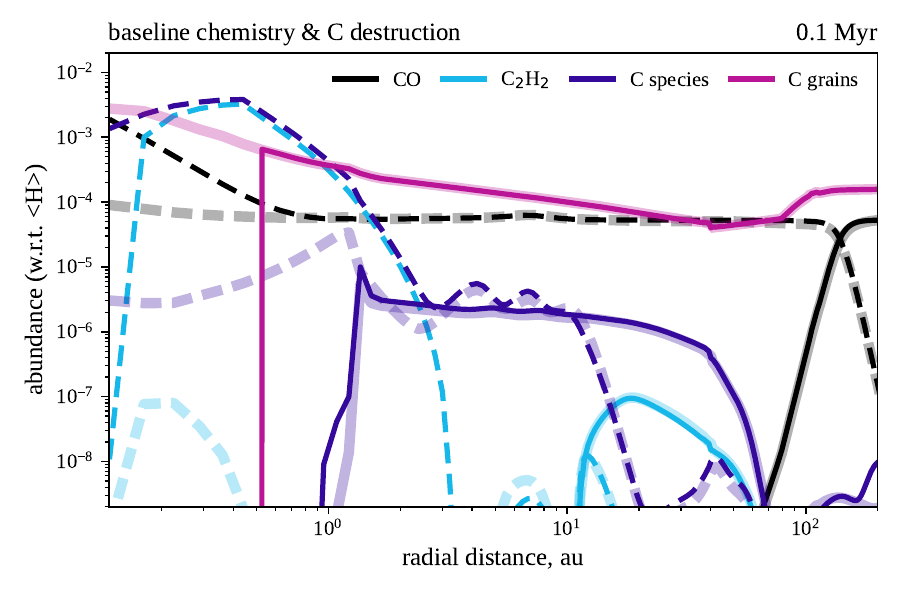}
\includegraphics[width=1\columnwidth]{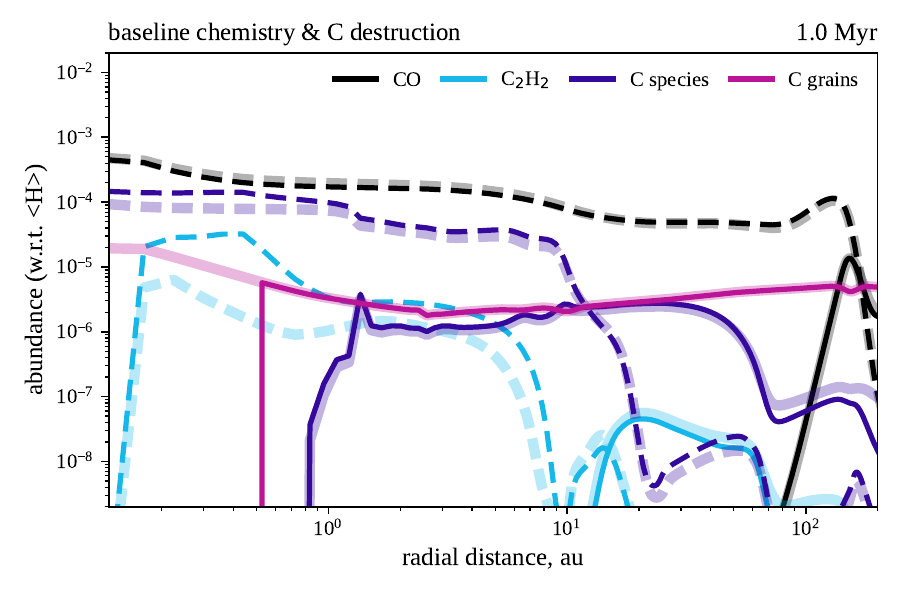}
    \caption{Radial distribution of abundances of some carbon-bearing species in the models with no dust trap and baseline C chemistry (with carbon grain destruction), at~0.1 and~1\,Myr. The thick transparent lines refer to the same model without carbon grain destruction. ``C~grains'' denotes the abundance of elemental carbon inside the dust grains. Acetylene \ce{C2H2} is included in the C~species.}
    \label{fig:c2o_carbon_grains_radial}
\end{figure}

As was recently shown by \citet{2025A&A...699A.227H}, the contribution from carbon grain destruction can increase the C/O ratio in the inner disc. To test the sensitivity of our results to carbon grain destruction, we add this process to the model, following the approach of \citet{2025A&A...699A.227H}. In our initial abundances, 60\% of all elementary carbon is hidden in the dust grains. In the models with carbon grain destruction, all this carbon is released instantaneously at temperatures $>350$\,K, forming acetylene \ce{C2H2}. The choice of a relatively low value of the critical temperature and the amount of carbon destroyed at this temperature (all the refractory carbon) should provide us with an upper limit on the expected effect of carbon grain destruction. However, some models include destruction of solid carbon at even lower temperatures, e.g., \citet{2025A&A...701A.194P}. For a young solar mass star, 350\,K  is achieved at $\approx0.5$\,au from the star, well inside the considered inner disc region.

\begin{figure}
\includegraphics[width=1\columnwidth]{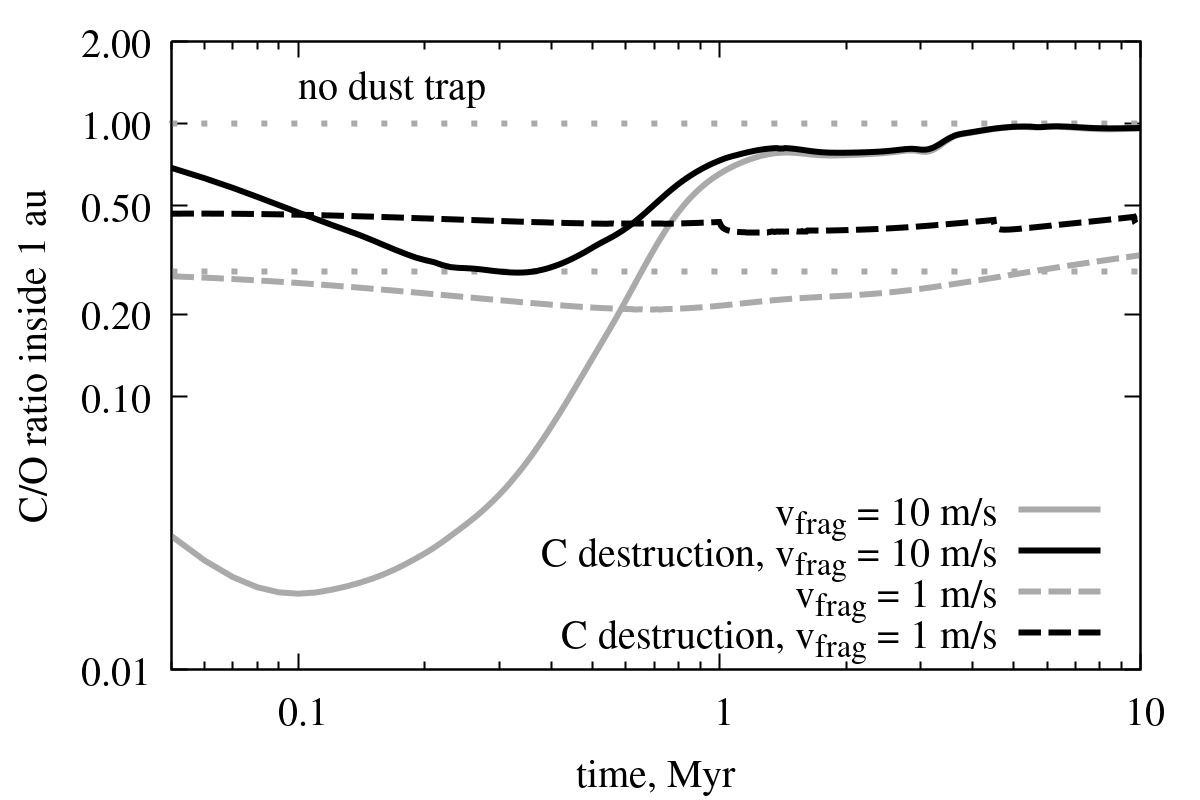}
    \caption{Gas-phase C/O ratios in the inner disc for baseline chemistry models with no dust trap, with and without carbon grain destruction.}
    \label{fig:c2o_carbon_grains}
\end{figure}

The choice of acetylene as the product is also arbitrary, as the process could release \ce{C+}, \ce{CH4}, or other more complex molecules. The choice of the product of carbon grain destruction could affect the composition of the carbon-bearing species forming in the inner disc, which should be considered in a separate study. However, this process is mostly relevant for the very inner disc ($<0.5$\,au), well inside the considered 1\,au region. The particular composition of the products should not significantly affect the total carbon and oxygen balance in the inner disc, as most of the products will not be able to escape this region through diffusion or viscous spreading. We do not expect the gas-phase C/O ratio in the inner region to be sensitive to the choice of the product of carbon grain destruction.

Figure~\ref{fig:c2o_carbon_grains_radial} shows the radial distribution of the carbon locked in the grains and released to the gas phase. The inclusion of carbon grain destruction releases acetylene inside 0.5\,au, followed by its outward diffusion and transformation to other species. Inside $\approx0.3$\,au it chemically transforms to CO and other carbon-bearing species (with \ce{CO2} and \ce{CH4} abundances insensitive to the increase in the acetylene abundance). At 0.1\,Myr, this additional acetylene is more abundant than CO in the inner disc, and contributes a lot to the carbon-oxygen balance. At 1\,Myr, the dust is already depleted by $1-2$ orders of magnitude, and the contribution from solid carbon is relatively small. \ce{C2H2} does spread outwards, but unlike in~\citet{2025A&A...699A.227H}, in our model it does not reach the location of the acetylene snowline before the dust reservoir is depleted, therefore, the released carbon is not conserved in the disc. One of the reasons is a slightly different sublimation temperature of \ce{C2H2}: we adopt the binding energy of $2090$\,K \citep{2004MNRAS.354.1133C,2017ApJ...844...71P}, which corresponds to the freeze-out at $\approx60$\,K, while \citet{2025A&A...699A.227H} assume that it freezes at 70\,K.

The main difference between our results and \citet{2025A&A...699A.227H} is largely due to our higher $v_{\rm frag}$, which increases the rate of dust transport and quickly depletes dust from the inner disc, weakening the effect of carbon release. The evolution of the inner disc C/O ratio in models with and without carbon grain destruction is shown in Figure~\ref{fig:c2o_carbon_grains}. Before $0.5-1$\,Myr, the contribution from the destroyed grains significantly elevates the C/O ratio inside 1\,au in the fast drift model ($v_{\rm frag}=10$\,m~s$^{-1}$). The amount of the excess carbon is comparable to the amount of volatiles, particularly water, brought from the outer disc by the radial drift. However, when the flux of pebbles is exhausted, the contribution of carbon grains becomes less significant. In the model with slow drift ($v_{\rm frag}=1$\,m~s$^{-1}$), the effect of carbon grain destruction is weaker at earlier times, but remains significant at $>1$\,Myr age, increasing the C/O ratio by a factor of $\approx2$. At 2\,Myr, the dust-to-gas mass ratio inside 1\,au is still around $10^{-2}$, much higher than the value of $10^{-4}$ in the fast drift case. The inclusion of a dust trap would further decrease the dust mass fraction, weakening the effect of carbon grain destruction.

We find that, in our fast drift models, carbon grain destruction is only efficient in enriching the gas in carbon at the earlier stages, while there is a significant amount of inwards drifting dust. It increases the minimum C/O in the inner disc achieved at a few kyr from $<0.05$ to $\approx0.2-0.3$.
After 1\,Myr its effect is minimal due to the loss of most dust in the inner disc. It does not help to increase C/O ratio in a few Myr old protoplanetary discs, assuming dust growth is efficient (high $v_{\rm frag}$, fast dust transport). In the case of a smaller average dust size, the effect is more significant, as e.g. in \citet{2025A&A...699A.227H}.

\section{Discussion}
\label{sec:discussion}

\subsection{Effects on the inner disc C/O due to dust drift, chemistry, and dust traps}
\label{sec:discussion_CtoO_effects}

JWST observations probe the C/O ratio of the inner discs in a range of objects from different star forming regions, with and without substructures, around very low to intermediate mass stars \citep[][]{2023ApJ...957L..22B,2023ApJ...947L...6G,2024ApJ...977..173C,2024PASP..136e4302H,2024ApJ...964...36R,2025AJ....170...67A,2025ApJ...991L..46C,2025ApJ...978L..30L,2026MNRAS.545f2056K}. To observationally test the predictions of our modelling, we need to look at discs at a population level across different ages. Our results suggest that radial drift will have the major effect on the inner disc elemental composition, with chemistry being important for discs on longer than Myr timescales.

The contribution of chemistry to the inner disc composition depends on whether dust transport is efficient or not. We consider a constant value of $\alpha=10^{-3}$, while lower values could be present in discs. At lower $\alpha$, dust can grow to larger sizes and drift faster, increasing the role of transport and leaving chemistry less time to change the composition of ices \citep{2019MNRAS.487.3998B}. At lower turbulence, the flux of pebbles is stronger but exhausts the dust source faster \citep{2024A&A...686L..17M,2025A&A...699A.227H}. Another uncertain parameter is the dust fragmentation velocity. At values lower than the adopted $10$\,m~s$^{-1}$, the growth will be restricted, weakening the transport. With less efficient radial drift, the enrichment of the inner disc with volatiles is expected to happen later and be weaker. The observed composition of the inner disc could inform us about the efficiency of radial drift: for example, observing low C/O in the inner discs of $\sim$Myr old objects could indicate slower transport.

The early chemical processing of methane has two main outcomes that significantly change the previous understanding of volatile transport with pebbles. The lack of carbon-rich gas in the outer disc prevents late carbon enrichment of the inner disc, meaning a maximum C/O of~1 at a few Myr age. The solid carbon is locked in a dust trap as well as water, meaning that inner discs of old discs with traps are expected to be poor in carbon, as well as in oxygen. Next, we expand on the implications of these two ideas and speculate on alternative ways to produce the observed high C/O ratios.

\subsubsection{Late carbon enrichment prevented by chemical processing}
\label{sec:carbon_enrichment_not}

Previous simulations show that high C/O ratio in the inner disc can be achieved at older ages due to viscous spreading of carbon-rich gases (CO and particularly \ce{CH4}). \citet{2017MNRAS.469.3994B} and \citet{2023A&A...677L...7M} showed that more volatile species with far out snowlines are less affected by radial drift and deliver their carbon at $>\mathrm{Myr}$ timescales. We find that chemical conversion of carbon-rich gas to carbon-rich organic ice transported by pebble drift leads to early delivery of carbon in smooth discs and ultimately prevents C/O reaching high values ($>1$). This conversion is driven by cosmic-ray ionisation, and for ionization rates of $\zeta>10^{-18}$~s$^{-1}$ the inner disc C/O ratio stays $\approx1$ at $>\mathrm{Myr}$ age, as CO remains the main carbon carrier, while \ce{CH4} is destroyed. JWST surveys of $>5~\mathrm{Myr}$ old discs in Upper Scorpius (PID 2970, PI: I. Pascucci) indeed confirm $\mathrm{C/O}<1$ in the inner discs, consistent with this prediction \citep{2026AJ....172..321X}.

Late carbon enrichment requires a high initial abundance of \ce{CH4} to achieve $\mathrm{C/O}>1$. The \ce{CH4} abundance we adopt from \citet{2023A&A...677L...7M} (50\% of CO abundance) is relatively high, as discussed by \citet{2025A&A...699A.227H}. In protostellar cores, the \ce{CH4} to \ce{CO} ratio is around 15\% \citep{2011ApJ...740..109O}, similar to the Solar System comets where it ranges from a few per cent to comparable amounts \citep{2011ARA&A..49..471M}. \citet{2025A&A...701A.239S} proposed a way to achieve higher methane abundances in the disc, based on the formation of \ce{CH4} from CO and further carbon enrichment at later stages derived from a phenomenological chemical model. We find that the inclusion of more comprehensive chemistry leads to efficient processing of \ce{CH4} into carbon-rich organic ices, thereby restricting the C/O ratios in the inner disc to values around~1, as discussed in Section~\ref{sec:ionisation_rate}. This highlights the importance of a more comprehensive treatment of chemistry, including a full chemical network with surface reactions.

Another uncertainty concerns the methane binding energy to water ice: as calculated by \citet{2022ApJS..262...17B}, it is below $1000$\,K, which is lower than that of CO. In most astrophysical models, including ours, a higher value of $\approx1100-1300$\,K is adopted based on TPD measurements \citep[e.g.,][]{1996ApJ...467..684A,2016A&A...595A..83E,2022ESC.....6..597M}. This could have a major effect on the timing of methane products delivery to the inner disc in case of weak chemistry ($\zeta\le10^{-18}$\,s$^{-1}$). With a lower binding energy, the C/O in the inner disc would grow more slowly, but without decreasing at later stages, as \ce{CH4} would be the last and most C-rich volatile to be delivered. At higher ionisation rates ($\zeta>10^{-17}$\,s$^{-1}$), methane destruction before 1\,Myr makes this difference less significant, making it unlikely that the C/O in the inner disc ever exceeds~1.

\subsubsection{Chemical composition in the presence of a dust trap}
\label{sec:discussion_trap_chemistry}

Dust traps were previously shown to be efficient in blocking oxygen delivery by trapping water \citep{2021ApJ...921...84K, 2023ApJ...954...66K,2024A&A...686L..17M,2024ApJ...977...21E}. Our results demonstrate that due to chemical processing, carbon is also efficiently trapped, which prevents C/O ratio increase in transition discs. 
Carbon-rich species originate from beyond the trap and thus never reach the inner disc, which also decreases its metallicity compared to smooth discs by an order of magnitude after $1-2$\,Myr. Discs with dust traps retain more dust and are typically brighter, implying the anticorrelation between dust flux and the C/H ratio in the inner disc.
However, if CRPHOT dissociation of ices is efficient, an additional chemical path appears for oxygen to escape the trap. In this case, the inner disc becomes oxygen-rich: it has super-solar metallicity and a low C/O ratio. Chemical reactions also provide a way to form \ce{CO} that can subsequently escape the trap, but with a lower efficiency (see Appendix~\ref{sec:O2_formation}).

Similar effects can be produced by leaky dust traps, which provide a way for dust grains and their ice mantles to escape the trap and reach the inner disc \citep{2023A&A...670L...5S,2025ApJ...988...94H,2025MNRAS.541.3101S}. As discussed by \citet{2026arXiv260411925H}, dust traps at $>5$\,au produce oxygen-rich inner discs with $\mathrm{C/O}<1$ without any chemical processing, if dust evolution is modelled with a full dust size distribution code accurately capturing coagulation and fragmentation. Our simulations therefore should underestimate the amount of carbon and oxygen escaping the trap with small grains produced in collisions. This process should be able to balance the overestimated oxygen production in our model to some degree. However, hydrocarbon ices would also be able to escape with leaking grains, adding carbon to the inner disc.

\subsubsection{Perspectives of $\mathrm{C/O}>1$ in the inner disc}
\label{sec:high_c2o_where}

JWST observations find $\mathrm{C/O}>1$ in the inner discs of a variety of protoplanetary discs. Carbon-rich chemistry is found to dominate around lower-mass stars \citep{2025A&A...699A.194A,2025A&A...702A.126G}, down to planetary-mass objects \citep{2025ApJ...991L..46C}. However, $\mathrm{C/O}>1$ was detected in discs around some solar-mass objects that have low accretion rates as well \citep{2024ApJ...977..173C, 2025arXiv251108816V}. If carbon-rich gas is destroyed by chemical processing as suggested here, alternative carbon enrichment mechanisms are necessary to explain the observations.

In the presence of active chemistry, achieving high C/O ratio values is not straightforward. Our modelling predicts the inner disc C/O typically around~1 for a~few~Myr old objects, ranging from $0.7-1.7$ depending on $\zeta$. A C/O noticeably above~1 can only be produced in low CR environments and when cosmic-ray-induced photo-dissociation of ices is not included, provided that the initial methane abundance is high. Such low CR environments could be a result of cosmic ray exclusion by the stellar wind \citep{2013ApJ...772....5C,2020MNRAS.499.2124R}, however locally-accelerated stellar cosmic rays may become important \citep{2021MNRAS.504.1519R}. 
Values marginally above~1 ($\approx1.1$) can also be achieved at around 1\,Myr for specific combinations of chemistry and ionisation rate (see Figure~\ref{fig:CR_basechem_fullchem}). Direct dissociation of CO by cosmic rays and \ce{O2} production on grains in dust traps  can further decrease the C/O down to $0.1-0.5$.

A way to increase the C/O ratio is the inclusion of carbon grain destruction. However we find that when transport is efficient, the reservoir of carbon solids is quickly depleted by drift anyway, and the C/O ratio only increases at early times as shown in Section~\ref{sec:carbon_destruction}. More significant amount of extra carbon could be released to the gas if dust accumulation were combined with variable temperature profile, e.g. due to luminosity outbursts, as \citet{2025A&A...699A.227H} also suggested. This mechanism might explain the solar-mass stars with unusually high carbon content, particularly if dust accumulated in a dust trap in the inner disc.

For low-mass stars, a more common mechanism should be at play to explain a systematically high $\mathrm{C/O}>1$. In this study, we only consider solar mass stars. A dust evolution model suitable for different stellar masses would be required for such simulations \citep[such as, e.g., \texttt{TriPod},][]{2024A&A...691A..45P}. Stellar mass is expected to affect mainly the timescales of the radial transport, with pebble drift bringing volatiles faster in discs around low-mass stars as the snowlines are closer \citep{2023A&A...677L...7M}.
Dust growth itself is determined by the combined effect of lower density, larger scale height and different temperature profile.
Overall, the radial transport of dust is found to be more efficient around lower-mass stars \citep{2013A&A...554A..95P,2023A&A...677L...7M}.

From chemical point of view, the timescales of methane processing in the midplane for low-mass stars should be similar to solar-mass stars. This means that our modelling would not predict C/O ratio above~1 in the discs of low-mass stars, contrasting the observations. Freeze-out based simulations of the chemical composition of photo-evaporating discs around low-mass stars also showed a C/O ratio below the typical observed values \citep{2025A&A...700A..67L}. The observed correlation of high carbon content with low accretion rate and low stellar mass might not be a consequence of radial drift or chemical processing in the outer disc, and an alternative explanation is required. The effects of chemistry on the composition of discs around low-mass stars should be investigated in a separate study.

\subsection{Scrutiny of the \ce{O2} production efficiency}
\label{sec:O2_uncertainty}

One of the substantial model-dependent results of our work is the cosmic ray induced production of \ce{O2} in the dust trap (see Appendix~\ref{sec:O2_formation}). This mechanism provides a way to release oxygen locked in the dust trap at a few Myr age, significantly affecting the inner disc C/O ratio. At the same time, it is unclear what fraction of radicals produced in the bulk of ice mantles can participate in surface reactions to trigger the \ce{O2} production.

Dissociation of water ice by CRPHOT in the full chemistry models is assumed to be happening in the whole of the ice mantle, not just in the upper layers, potentially significantly overestimating the involvement of the produced \ce{OH} radicals in surface reactions. The effect of \ce{O2} formation is an extreme case happening in most favourable conditions, as we show in Section~\ref{sec:CRPHOT_ices_trap} and Figure~\ref{fig:CRPHOT_mantle_coefficient}, and is sensitive to the fraction of the mantle affected by the CRPHOT dissociation of ices. The use of a three-phase or multi-phase ice model may mitigate \ce{O2} formation as well. Even if \ce{O2} forms in the ice mantle, a reasonable fraction of it is expected to remain trapped deep in the bulk, while only the surface-formed \ce{O2} would have a chance to escape \citep[e.g.][]{2016MNRAS.462S..99T}.

Another uncertainty concerns the rate coefficients of cosmic-ray–induced photo-reactions, adopted from \citet{1989ApJ...347..289G}, which assume a dust size distribution of small interstellar grains.
However, dust growth in the disc midplane should reduce photon absorption, leaving a stronger UV field and increasing reaction rates. A two–order-of-magnitude increase in maximum grain size corresponds to an order-of-magnitude increase in average grain size and hence in CRPHOT reaction rates, potentially altering ice compositions \citep{2021INASR...6..122Z}. This would enhance water-ice dissociation and accelerate \ce{O2} formation. On the other hand, the dust-to-gas mass ratio in the dust trap is also much higher than was assumed by \citet{1989ApJ...347..289G} ($\sim0.5$ compared to the standard 0.01). The increased dust abundance should lead to more efficient adsorption of CRPHOT and thus lower ionisation rates. To some degree, the effects of dust size and mass fraction should compensate each other, but their importance should be separately investigated.

\subsection{Observational limits on the \ce{O2} formation}
\label{sec:O2_observation_discussion}

\begin{figure*}
\includegraphics[width=0.66\columnwidth]{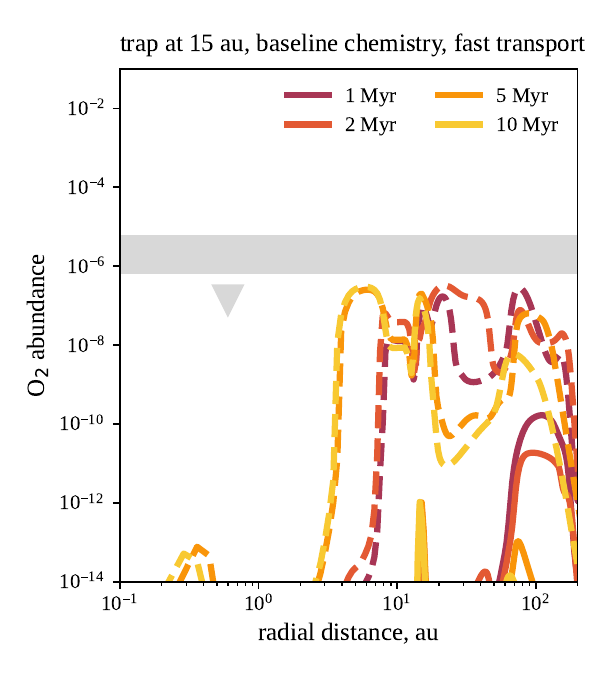}
\includegraphics[width=0.66\columnwidth]{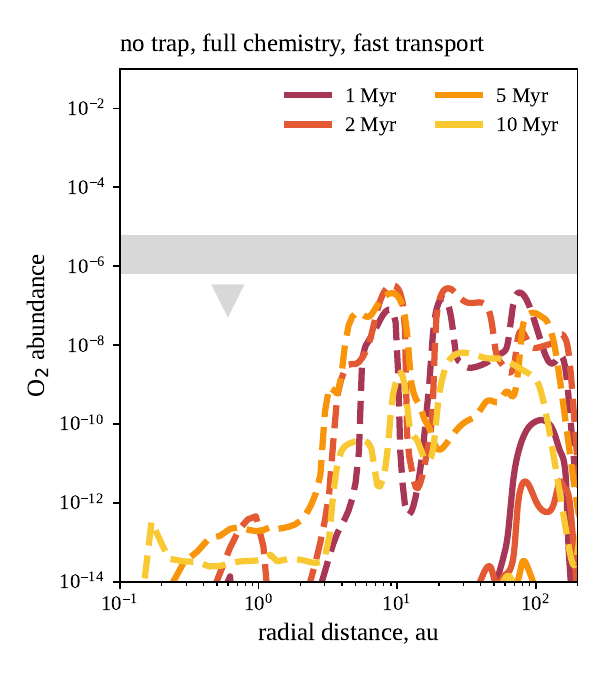}
\includegraphics[width=0.66\columnwidth]{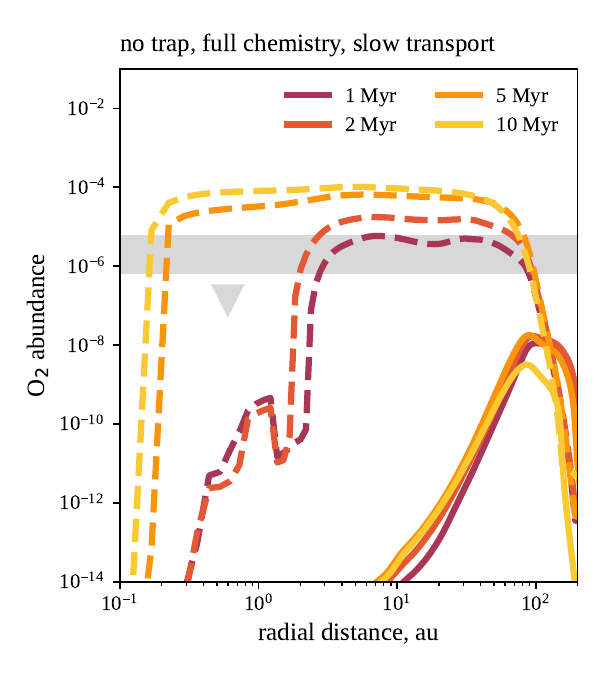}
\includegraphics[width=0.66\columnwidth]{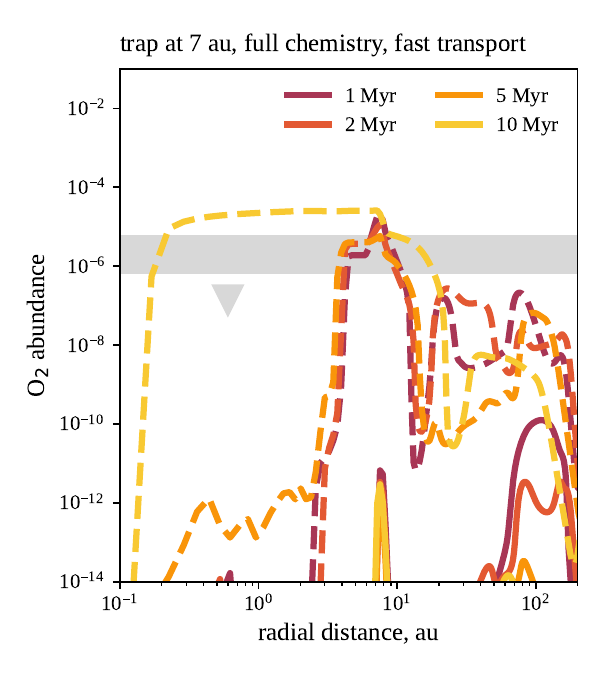}
\includegraphics[width=0.66\columnwidth]{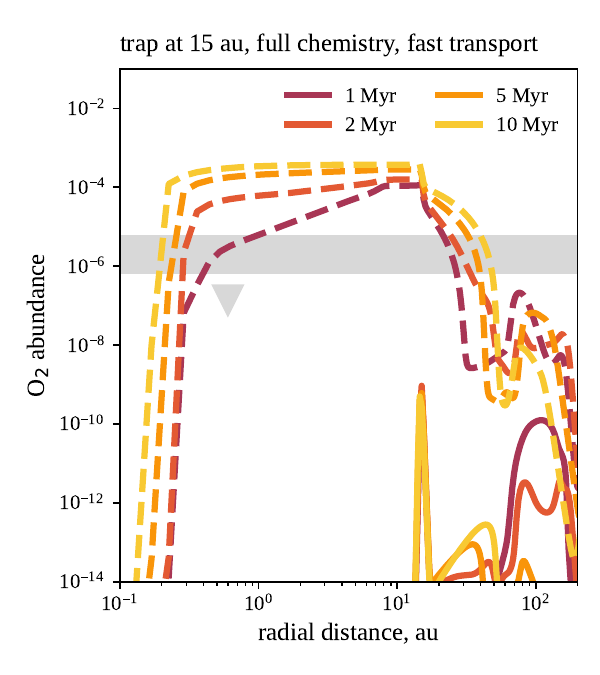}
\includegraphics[width=0.66\columnwidth]{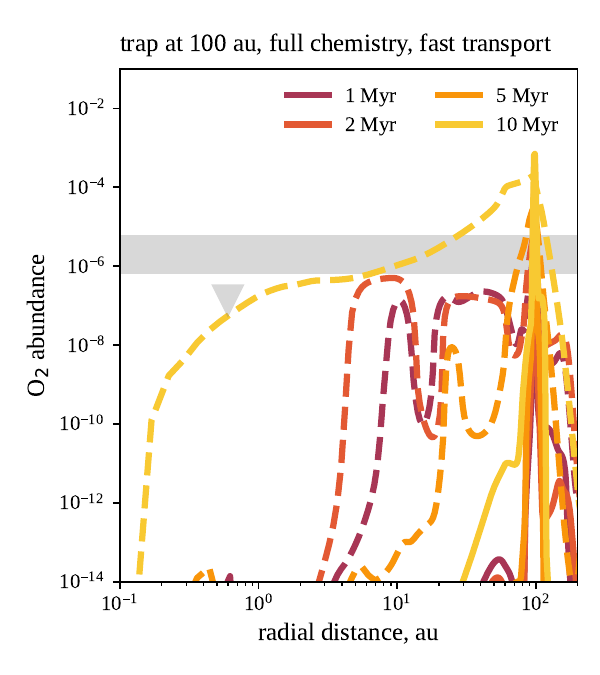}
    \caption{Relative abundances of \ce{O2} in the gas and in the ice in models with different locations of the dust trap and different chemical regimes. The shaded area marks the observational upper limit on \ce{O2} abundance in TW~Hya \citep{2023ApJ...956..135W}. }
    \label{fig:trap_radials_O2}
\end{figure*}

If molecular oxygen is indeed efficiently produced in the dust trap, it could be observable. At the moment, ALMA observations aimed at molecular oxygen have not yielded any detections. For the disc around TW~Hya, \citet{2023ApJ...956..135W} used the emission of the isotopologue $^{16}$O$^{18}$O to determine an upper limit on the \ce{O2} abundance of $(6.4-62)\times10^{-7}$.
\ce{O2} with an abundance of a few percent relative to water was also detected in two Solar system comets, 67P~/C-G  \citep{2015Natur.526..678B} and 1P~/Halley \citep{2015ApJ...815L..11R}. The  origin of the \ce{O2} ice is uncertain, but astrochemical modelling suggests that it is likely primordial \citep{2016MNRAS.462S..99T,2019A&A...621A..75E}.

Figure~\ref{fig:trap_radials_O2} shows the radial distribution of \ce{O2} in our models in comparison with the observational limit from \citet{2023ApJ...956..135W}. 
With the baseline chemistry, the \ce{O2} abundance in the gas stays at $10^{-10}-10^{-7}$ at all times. With full chemistry and no traps, \ce{O2} abundance can exceed the observational limit only in case of slow transport, which is inconsistent with the $>10^{-9}$\,$M_{\odot}$~yr$^{-1}$ accretion rate of  TW~Hya \citep{2023ApJ...956..102H}. In all models with full chemistry and a dust trap, the \ce{O2} abundance exceeds $10^{-5}$ at some point in time, depending on the trap location. The high \ce{O2} abundance reached by 5\,Myr in models with a dust trap at intermediate distances (tens of au) is comparable with that of CO and covers a large enough disc region to be potentially detectable at $\sim40$\,au resolution. 

The non-detection of molecular oxygen in TW~Hya may point to a limited contribution of \ce{O2} formation, either due to inefficient CRPHOT-induced ice dissociation or a low ionisation rate. Observational constraints from \ce{HCO+} and \ce{N2H+} emission do indicate a CR ionisation rate below $10^{-19}$\,s$^{-1}$ in TW~Hya~\citep{2015ApJ...799..204C}. Methanol observations in TW~Hya confirm that grain surface chemistry plays an important role in setting the abundances of the observed species there, but the efficiency at which ices are photodissociated is overestimated in the modelling \citep{2025arXiv251004106I}. Alternatively, the rings of TW~Hya might not provide high enough dust-to-gas ratio ($\sim0.1$) for long enough time ($\sim1$\,Myr). Observational limits rule out efficient ice-phase production of molecular oxygen in dust traps in TW~Hya.

An older transition disc with a prominent dust trap at tens of au distance would be a perfect candidate to look for molecular oxygen. Efficient \ce{O2} formation would lead to super-solar metallicity (see second right panel in Figure~\ref{fig:volatiles_1au_c2o}) and high water abundance in the inner disc. Without this mechanism, at a few Myr age the inner disc should be poor in both carbon and oxygen, unless the dust trap is leaky. 
There are examples of bright water emission and a low C/O ratio in large discs or older transition discs~\citep{2025arXiv250104587G}. Particularly, PDS~70 and SY~Cha (around~5 and 3~Myr old, respectively) each possess a bright ring in dust emission, a very deep, extended gap, at the same time demonstrating strong emission of water brighter than that of carbon species. Another example is Sz~98 \citep{2023A&A...679A.117G}, which has no prominent dust rings, but a discontinuity in its radial emission profile that might indicate the presence of a dust trap~\citep{2018ApJ...859...21A}.

Transition discs with oxygen rich inner discs, such as PDS~70 or SY~Cha could be promising targets for a potential search for \ce{O2} emission with ALMA. 
SY~Cha has accretion rate of $3.89 \times 10^{-10}$\,$M_{\odot}$~yr$^{-1}$ and molecular emission from inside the cavity  \citep{2017A&A...604A.127M,2023PASJ...75..424O,2025ApJ...984L...6T}. Its inner disc is rich in both oxygen and carbon bearing molecules \citep{2024ApJ...962....8S}. PDS~70, an older object and a host of the first planets detected in a protoplanetary discs, also has weak accretion rate of $\approx 2 \times 10^{-10}$\,$M_{\odot}$~yr$^{-1}$. ALMA observations show high C/O ratio $\approx1$ in the outer disc around the dust ring and weak line emission from within the cavity~\citep{2024A&A...689A..65R}. However, JWST~MIRI data show water emission and no hydrocarbons in the inner disc \citep{2023Natur.620..516P}, indicating a lower C/O ratio. These observations do not necessarily require \ce{O2} production to explain, because a leaky dust trap could potentially enrich inner disc with water \citep{2025MNRAS.541.3101S}, but the observed composition leaves room for possible \ce{O2} production in the dust trap.  Hot water is also detected in the inner disc of AS~209 which has large prominent rings \citep{2024ApJ...964...36R}. However, in some transition discs hydrocarbons dominate the spectrum, suggesting a  high C/O ratio \citep{2025arXiv251108816V}. The diversity of transition disc composition and possible reasons for differences in the inner disc composition are discussed in Section~4 of \citet{2025arXiv251108816V}. Nevertheless, their high concentration of dust and thus potentially of water ice makes them more favourable candidates for \ce{O2} detection than TW~Hya.

Potentially oxygen-rich gas is observed also in some Herbig~Ae/Be stars. ALMA observations show water emission just inside  the dust trap in the disc around HD~100546 \citep{2025arXiv251206439R}. In this object, the emission might be coming from inside of the water snowline, which is located at a similar distance as the trap ($\approx20$\,au). In the disc around HD~169142, the C/O ratio inside $30-60$\,au is found to be approximately solar, rising above 0.5 at larger distances \citep{2024MNRAS.534.3576K}. The C/O ratio increases approximately between the inner and outer dust rings, which might be connected with production of \ce{O2} in the dust traps. In Herbig stars,  their higher mass and luminosity will have two opposite effects on the radial drift: snowlines further out will lead to slower transport in the gas after volatiles sublimate, causing later enrichment, while a larger scale height would make the drift velocity higher. Dedicated chemical modelling is required to predict their C/O ratios.

\section{Conclusions}
\label{sec:conclusions}

We simulated the chemical evolution of a protoplanetary disc midplane to analyse the elemental composition of the gas in the inner disc regions of solar-type stars. We combine comprehensive chemical modelling including complex surface chemistry with two-population dust growth parametrisation and radial transport in a viscously evolving disc. We consider models with different transport regimes, with and without dust traps, and with different trap locations. We analyse chemical evolution with and without the contribution of ice-phase dissociation by cosmic ray induced photons and at different levels of cosmic ray ionisation rate.

We find that the C/O ratio in the inner disc regions is affected by chemical reactions at times longer than 1\,Myr. Unlike freeze-out only models, the models with chemistry predict a C/O ratio around and below~1 for a few Myr old discs. We focus on the models with relatively fast transport, characterised by $v_{\rm frag}=10$\,m~s$^{-1}$ and $\alpha=10^{-3}$, and highlight that in discs with less efficient radial drift the C/O ratio of the inner disc hardly evolves from the initial over 10\,Myr. The key parameters that govern the contribution of chemistry in our modelling are the incident cosmic ray ionisation rate, the efficiency of cosmic ray induced photo-reactions in the ice phase, and the presence and the location of a dust trap accumulating water ice. We find that with a high fragmentation velocity, the inclusion of carbon grain destruction only affects the C/O ratio in the inner disc during 1\,Myr, increasing it from below $0.1$ to $0.2-0.3$.

The main effect of chemistry is that it releases carbon from methane, causing earlier delivery of carbon to the inner disc because \ce{CH4} is converted to less volatile carbon-bearing molecules. This process is sensitive to the cosmic ray ionisation rate, and substantially affects the C/O ratio at $\zeta\gtrsim10^{-17}$\,s$^{-1}$. As a result, \ce{CH4} is no longer the main carbon carrier at Myr timescales. Instead, the C/O ratio is primarily determined by CO and \ce{CH3OH} formed from CO on the dust near the CO snowline after $\approx3$\,Myr. This sets an upper limit of $\approx1$ on the C/O in the inner regions after 3\,Myrs.

The inner disc can be significantly enriched with oxygen due to \ce{O2} formation in the dust trap, if cosmic ray dissociation of ices is efficient. This mechanism allows to release oxygen locked in the dust trap transforming it to more volatile \ce{O2}, if at least around 10\% of the dissociation products in the ice mantle take part in surface chemistry. Similarly, CO gas is produced from the destruction of carbon-rich ices.
Molecular oxygen has not yet been observed in protoplanetary discs, and the available upper limits are inconsistent with the predicted abundances for the most optimistic cases. The use of a three-phase or multi-phase ice model may reduce the production of \ce{O2} and \ce{CO} compared with the two-phase models used here, because even if formed in the ice, they may stay trapped within the water ice matrix. We expect that this mechanism should be most efficient in transition discs with highly concentrated dust rings, leading to water-rich inner discs, and suggest that \ce{O2} detection in such objects can be a signature of efficient cosmic-ray-induced photo-dissociation of ice.

\section*{Acknowledgements}
We thank the reviewer for their valuable suggestions, which helped shape and better present our results.
We thank I. Cleeves for productive discussions about surface chemistry. T. Molyarova and R.A.~Booth were supported by the Royal Society, award numbers URF\textbackslash R1\textbackslash 211799 and RF\textbackslash ERE\textbackslash 231082.  R.A.~Booth also acknowledges financial support from UKRI via grant number UKRI1191.
C.~Walsh~acknowledges financial support from the Science and Technology Facilities Council and UK Research and Innovation (grant numbers ST/X001016/1 and MR/Z00019X/1). This work was undertaken on the Aire~HPC system at the University of Leeds, UK. We thank V.~Akimkin and V.~Ivashkin for providing supplementary computational resources.

\section*{Data Availability}

Simulation results and data behind the figures will be shared upon request.

\bibliographystyle{mnras}
\bibliography{example} 



\appendix

\section{The C/O ratios in discs with dust traps}
\label{sec:c2o_with_trap}

\begin{figure*}
\includegraphics[width=0.97\columnwidth]{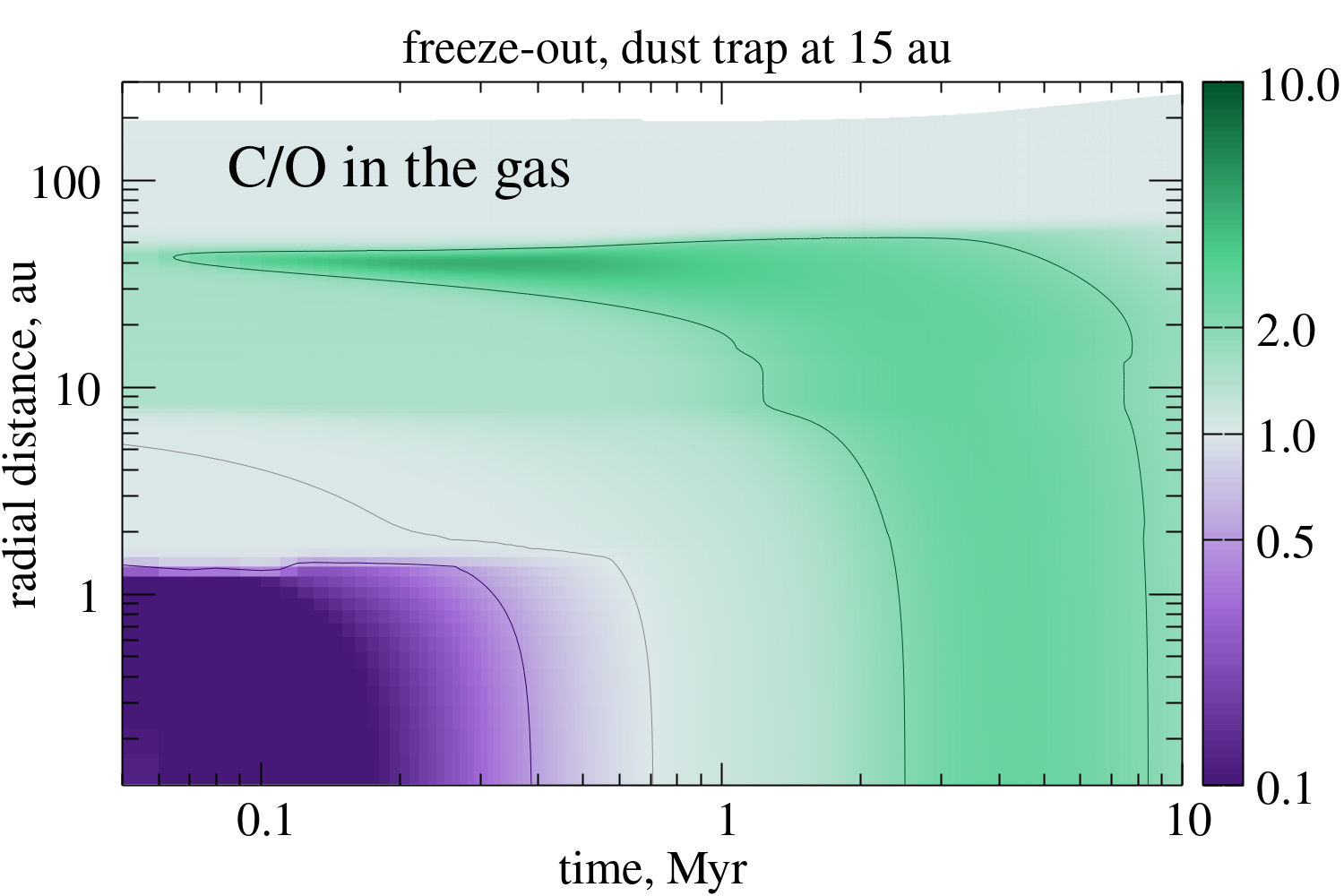}
\includegraphics[width=0.97\columnwidth]{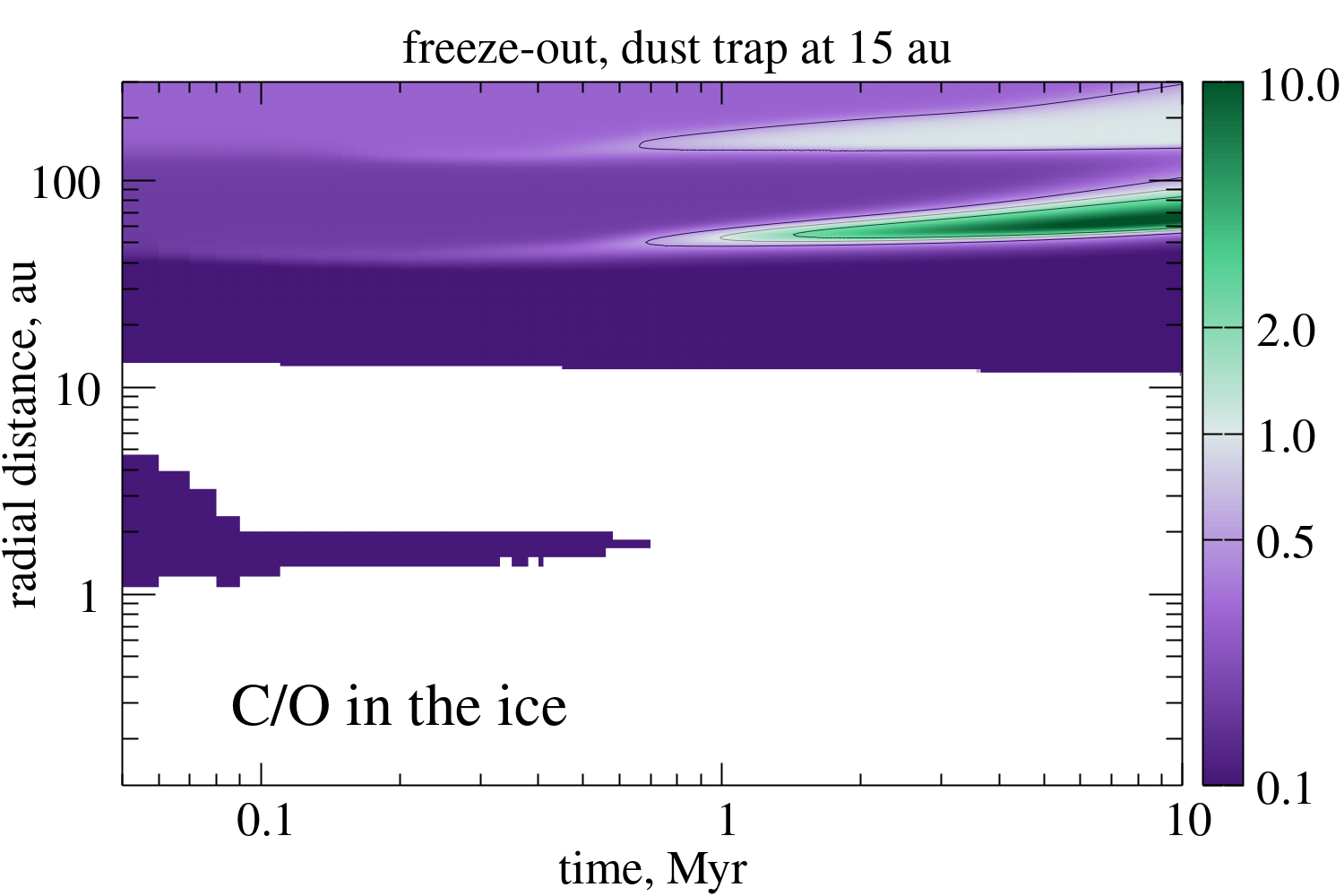}
\includegraphics[width=0.97\columnwidth]{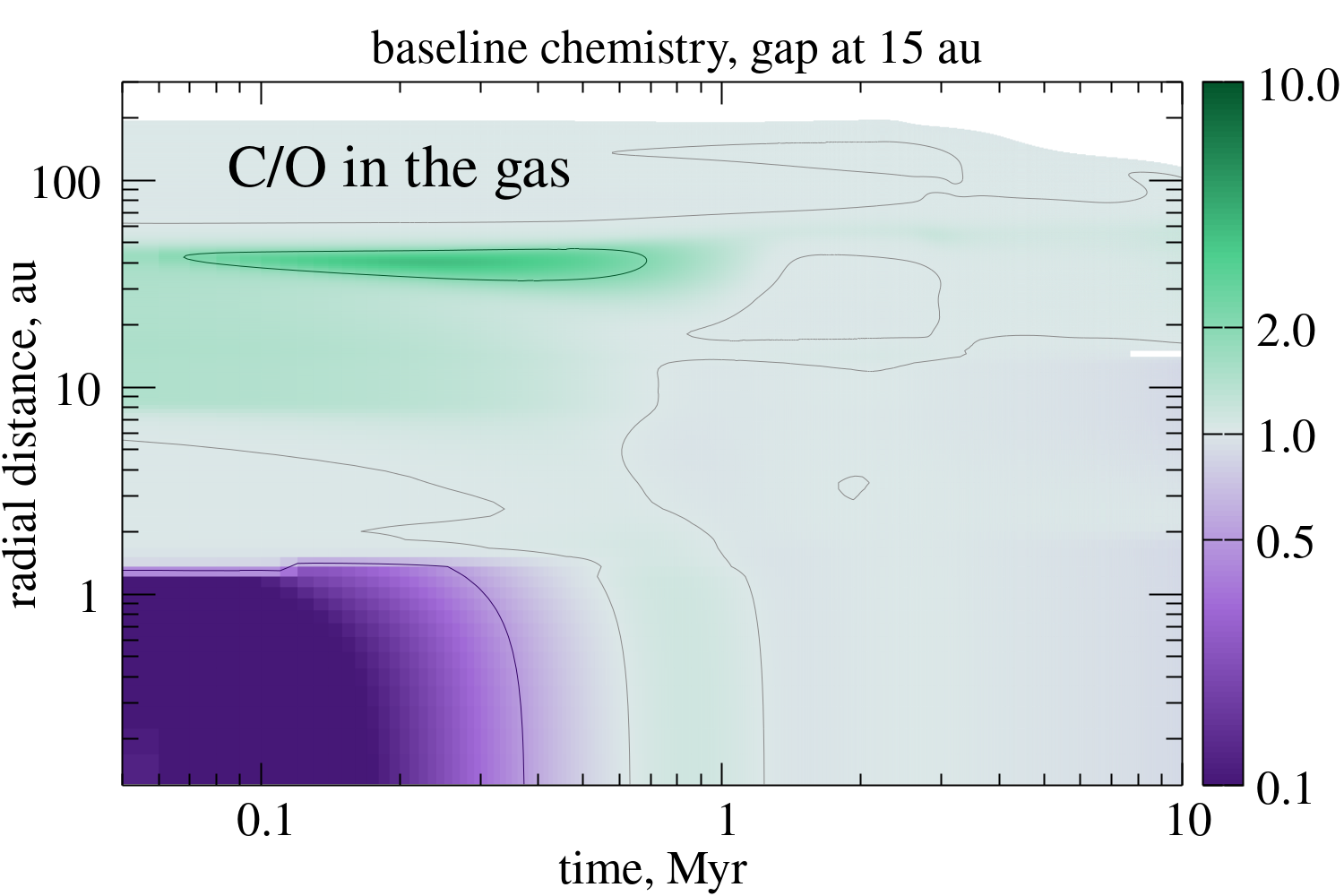}
\includegraphics[width=0.97\columnwidth]{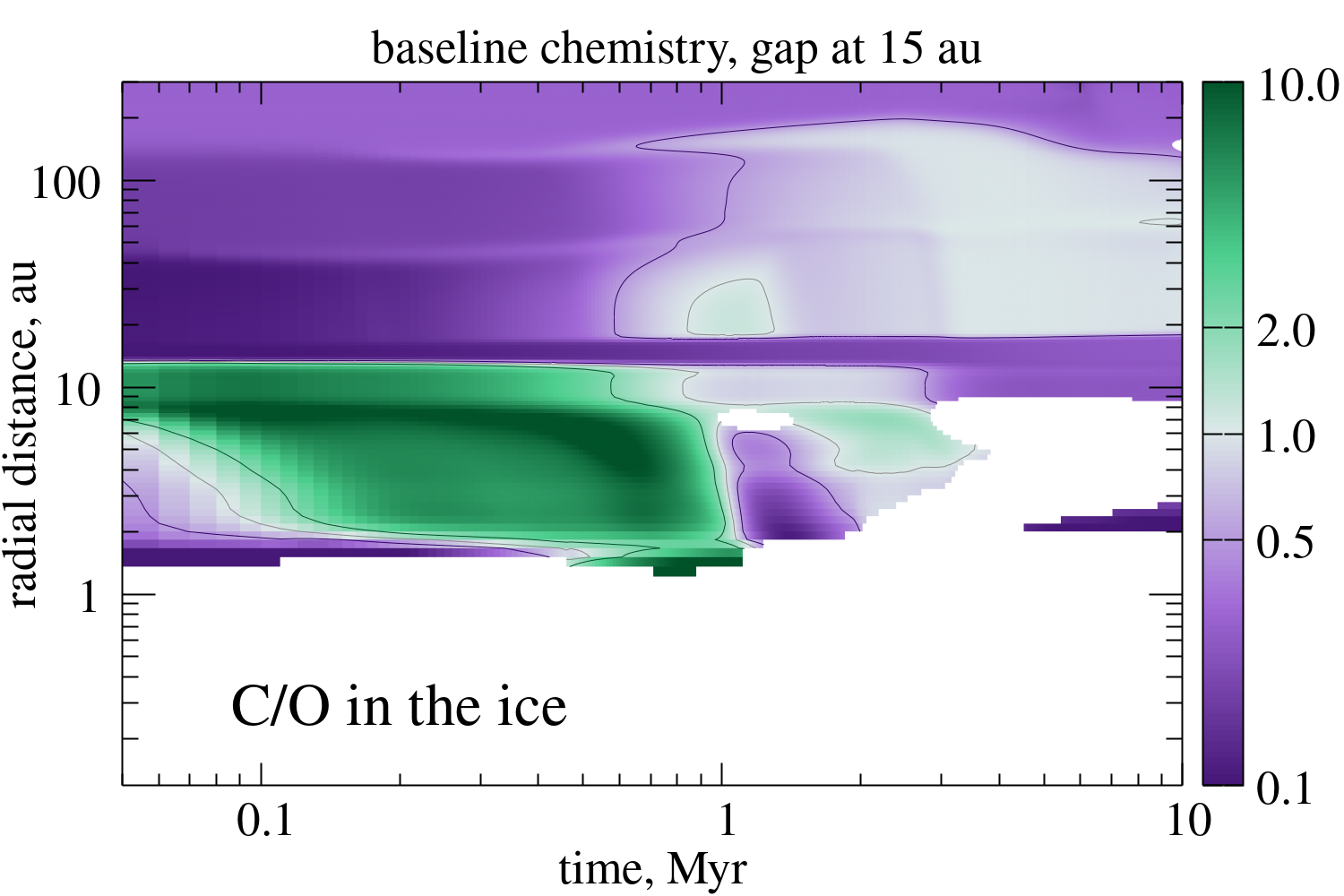}
\includegraphics[width=0.97\columnwidth]{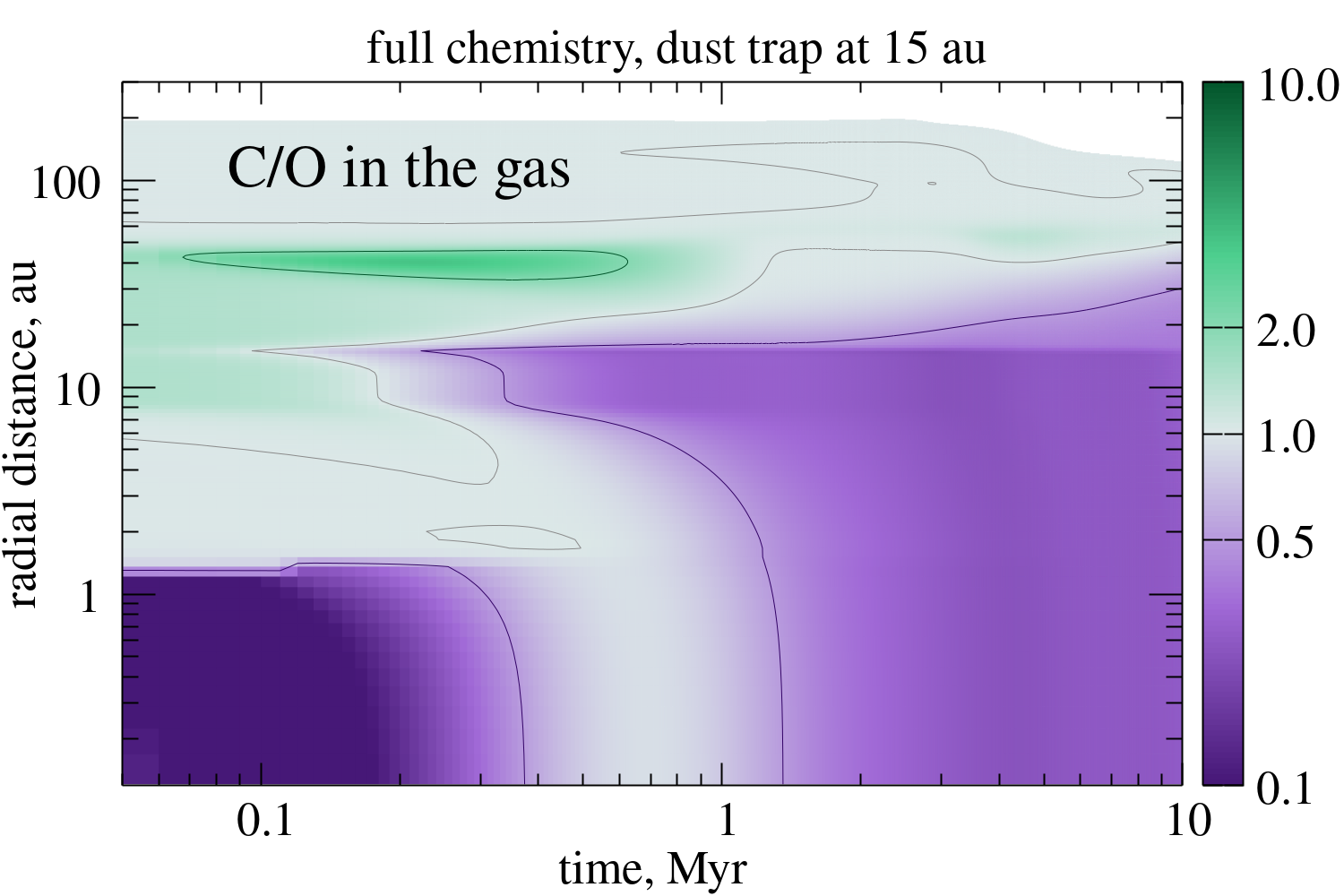}
\includegraphics[width=0.97\columnwidth]{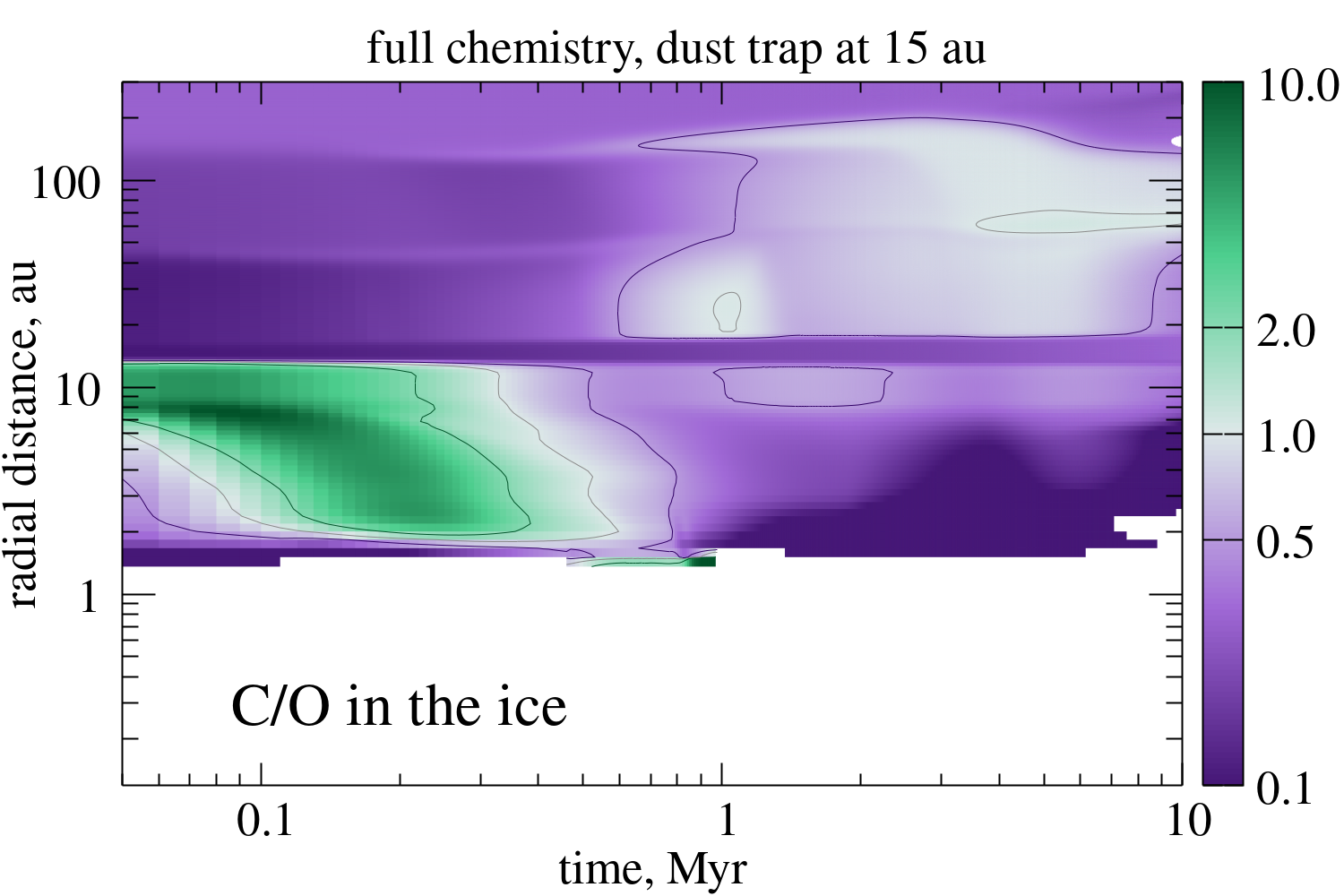}
    \caption{Evolution of the radial distributions of C/O ratios in the gas (left) and in the ice (right) in models with a dust trap at 15\,au. Top to bottom: freeze-out, baseline chemistry and full chemistry models. The regions where the total abundance of C and O in gas or ice is lower than 0.001 of their initial total abundance are shown in white. The contours mark the C/O ratio values of 0.5, 1 and 2.}
    \label{fig:c2o_evolution_baseline_gap}
\end{figure*}

Figure~\ref{fig:c2o_evolution_baseline_gap} summarises the evolution of gas- and ice-phase C/O ratios in the models with a dust trap at 15\,au. The region interior to the dust trap becomes depleted in ices as the dust drift clears it of solids. In the full chemistry model, formation of \ce{O2}, \ce{CO} and other species in the trap allows volatiles to escape the trap and leads to non-negligible amount of ice interior to the trap.

\section{Ice chemistry in the dust trap}
\label{sec:O2_formation}

In full chemistry models with dust traps, the disc composition at $\gtrsim$Myr timescales is qualitatively different from baseline chemistry models. Dissociation of ices by cosmic ray-induced photons triggers the chemical process that allow  oxygen to escape from the trap and leads to oxygen-rich inner disc with low gas-phase C/O ratio and high metallicity. Here we show the reaction routes responsible for this effect.

\ce{O2} formation in the dust trap starts with dissociation of water ice by cosmic ray induced photons:
\begin{equation}
    \ce{gH_2O ->[CRPHOT] gOH + gH}.
\end{equation}
Hydroxyl radicals in the ices on dust grain surfaces, \ce{gOH}, are involved in multiple reaction cycles (note that here we mark the species on dust grain surface with a letter g, for grain). Reacting with atomic hydrogen, they can form water ice again, in a barrier-free reaction \ce{gOH ->[gH] gH2O}. Multiple reaction routes lead to the formation of \ce{gO2}, following the release of atomic oxygen \ce{gO}:
\begin{equation}
\begin{aligned}
    \ce{gOH ->[gOH] gH_2O + gO}, \hspace{6pt} & \ce{gO ->[gO] gO_2} \\
                                              &  \ce{gO ->[gOH] gHO_2 ->[gO] gO_2 + gOH} \\
                                              &  \ce{gO ->[gHS] gSO ->[gO] gSO_2 ->[gH] gO_2} 
\end{aligned}
\end{equation}

\begin{figure}
\includegraphics[width=\columnwidth]{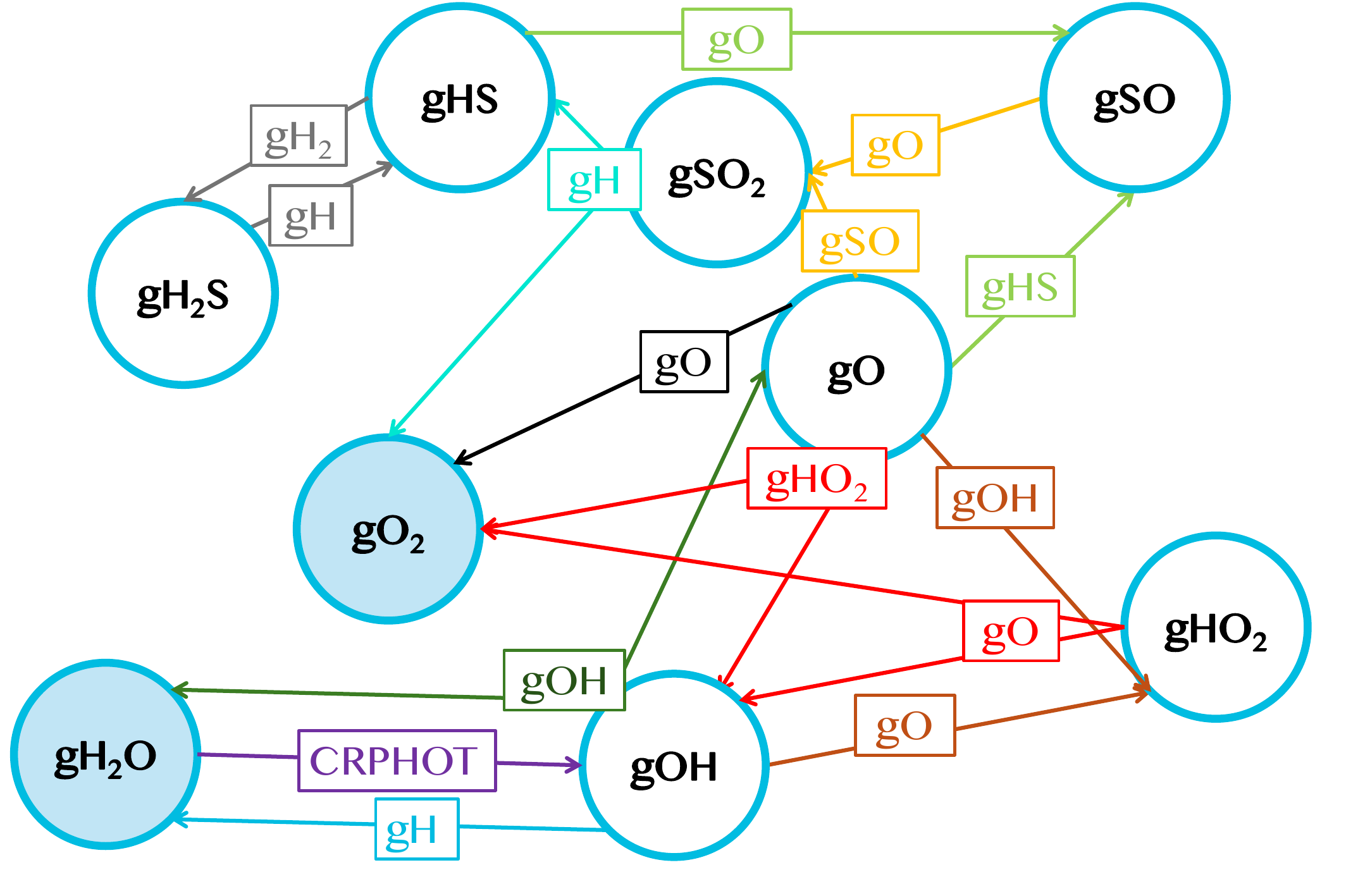}
    \caption{Main formation routes of \ce{O2} on dust surface. Most reactions with atomic hydrogen are excluded for clarity.}
    \label{fig:o2_form_diagram}
\end{figure}

These reactions include direct formation of \ce{gO2} from \ce{gO}, a pathway through the hydroperoxyl radical \ce{gHO2} and a cycle catalysed by sulphur-bearing species. These reactions are summarised in Figure~\ref{fig:o2_form_diagram}. Sulphur as a possible catalyst of \ce{O2} production in a process involving SO was described in \citet{2018A&A...613A..14E}. However, this process is only efficient at $>1$\,Myr time scales. Among these formation paths, the \ce{gHO2}/\ce{gOH} route is the most efficient. 

Apart from molecular oxygen, the dust trap stimulates the formation of \ce{CO} on dust grains. There is an ensemble of reactions that produce \ce{gCO}, one of which is directly linked to \ce{O2} formation:
\begin{equation}\label{eq:CO_form}
    \ce{gO_2 ->[gC] gCO + gO}.
\end{equation}
In the presence of plenty of \ce{gO2}, gCO will also inevitably form, because there is carbon provided by cosmic ray dissociation of complex hydrocarbons on dust surfaces. These hydrocarbons, such as \ce{gCH2CHCCH}, are formed in the outer disc regions and accumulate in the trap. Their formation is also triggered by methane destruction and subsequent reactions in the gas and in the ice. Like \ce{O2}, the formed CO sublimates and escapes the dust trap. Some of it is converted to \ce{CO2} on the way to the inner disc. In addition to Eq.~\eqref{eq:CO_form}, there are reaction routes that produce CO in the ice either through gHCO, gOCN, or sulphur bearing species such as gOCS and gHS. However, before sublimating to the gas phase, a significant fraction of the produced gCO also reacts with the available gOH to form \ce{gCO$_2$}. This restricts the efficiency of this channel, and helps to lock the \ce{CO2} in the trap. Only a fraction of the CO produced in the trap is able to escape to the gas phase.

Production of molecular oxygen could be overturned by the counteracting process of water formation. Water ice can form on dust surface through step-by-step hydrogenation of molecular oxygen, experimentally studied by \citet{2010PCCP...1212077C}:
\begin{equation}\label{eq:O2_H}
\begin{aligned}
    \ce{gO_2 ->[gH] gHO_2}, \hspace{6pt} & \ce{gHO_2 ->[gH]  gO + gH_2O} \\
                                         & \ce{gHO_2 ->[gH]  gOH ->[gH] gH_2O }.
\end{aligned}
\end{equation}
These processes depend strongly on the availability of H~atoms on the grain surface.  However, atomic H has a low binding energy (650\,K in the model) and desorbs rapidly at 53\,K, where the trap is located. This makes the reverse process relatively inefficient. Essentially, the \ce{O2} formation routes are more efficient because they do not rely on hydrogenation, particularly in the hydroperoxyl branch. For the sulphur catalytic cycle, hydrogenation is necessary to produce gHS, but this species can be formed in the outer disc regions and delivered to the trap.
Conversely, the key formation routes for \ce{O2} involve species with higher binding energies. The key reaction route with hydroperoxyl only includes species with binding energies  higher than that of atomic hydrogen: {gO (1\,660\,K), gOH (3\,210\,K) and \ce{gHO2} (1\,510\,K) \citep{2014ApJ...788...50H,2014PCCP...16.3493H,2017ApJ...844...71P}, which allows them to stay on the dust surface and form molecular oxygen faster than water.

Molecular oxygen formation from water should only be efficient within a certain temperature range depending on the locations of the snowlines of the involved components. The temperature should be high enough to prevent significant deposition of atomic hydrogen, but low enough to still avoid thermal desorption of hydroperoxyl and atomic oxygen. However even within their snowlines, species can briefly adsorb to the grain surface and participate in the surface reactions, because the timescale of diffusion and reaction can be shorter than the desorption timescales \citep[shown experimentally, e.g.][]{2022A&A...666A..35T}.  The highest temperature for the O and \ce{HO2} staying on the dust surface is around 70\,K, reached at $\approx7$\,au. The lowest temperature for \ce{O2} formation should be determined by the efficiency of water formation, and consequently the amount of atomic hydrogen, with unclear temperature range. As a hard limit, if any \ce{O2} is formed  below the \ce{O2} sublimation temperature $\approx20$\,K, it will not escape to the gas phase.

The surface chemistry of water and molecular oxygen at lower temperatures ($T\lesssim30$\,K) was experimentally investigated in multiple studies \citep{2008ApJ...686.1474I,2010PCCP...1212077C,2012ApJ...756...98J,2025ApJ...985..254S}. It was shown by \citet{2008ApJ...686.1474I} that water ice can be efficiently formed from \ce{O2} ice bombarded with hydrogen at 10-28\,K. Water production from O, \ce{O2} and \ce{O3} through multiple routes was investigated by \citet{2010PCCP...1212077C} at 15-25\,K. Laboratory data show the presence of intermediate products of these formation routes: \ce{H2O2}, \ce{O3}, \ce{HO2}. These species and all water formation routes are included in the presently used network. In our simulations, cosmic ray ionisation of ices provides an additional source of water destruction, increasing OH abundance and enhancing the reactions with hydroxyl, which shifts the balance to \ce{O2} production. We also consider higher temperatures, which correspond to lower H~abundances, further restricting water production, as well as more efficient \ce{O2} thermal desorption, which obstructs surface formation of water. The combination of relatively high temperature ($\approx50$\,K) and efficient dissociation of water ice are the key factors leading to efficient \ce{O2} production in the dust trap. There is also the assumption of a two-phase model: in reality, radicals/atoms formed in the bulk will have slower rates of diffusion which could push the reformation of water to be favoured over production of molecular oxygen.

The binding energies of H and O are among the key model parameters in the surface chemistry of water and molecular oxygen \citep{2015ApJ...801..120H}. They are relatively well constrained by the laboratory experiments \citep{2017SSRv..212....1C}, but some uncertainties could arise from the dust material and the type of the bond to it (chemisorption or physisorption). The gH abundance should also be sensitive to the \ce{H2} production mechanism \citep[see, e.g.,][]{2010MNRAS.406L..11C,2017MolAs...9....1W}. We consider models with higher H binding energy and a different~\ce{H2} formation mechanism in and find that \ce{O2} production is robust to these parameters. These models in comparison with the fiducial full chemistry model are shown in Figure~\ref{fig:H2_form}. The increase of the \ce{gH} abundance is not enough to surpass the \ce{O2} formation and has very little effect on the inner disc composition. In the model with Langmuir-Hinshelwood formation of \ce{H2} (as opposed to Eley-Rideal mechanism in the rest of the models), the amount of oxygen escaping the trap and reaching the inner disc also stays approximately the same, but the \ce{O2} is more efficiently converted to water on its way to the inner disc. The \ce{O2} formation mechanism is robust to the parameters of surface chemistry of hydrogen.

\begin{figure}
\includegraphics[width=\columnwidth]{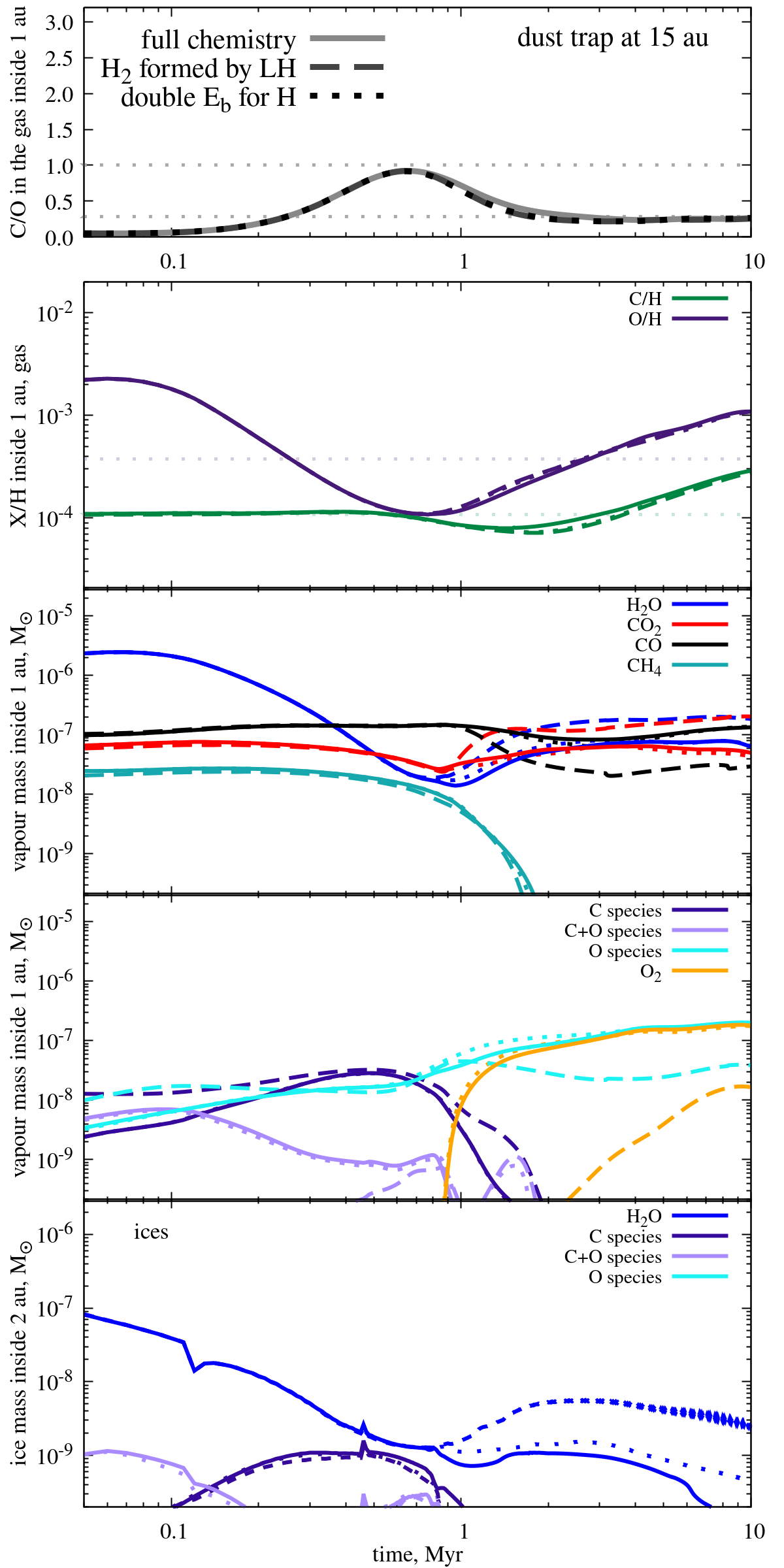}
    \caption{Evolution of integrated masses of main gas-phase volatiles and C/O ratios in the inner disc in the models with full chemistry, different H binding energy and different \ce{H2} formation mechanism. \ce{O2} is also included in the O~species.}
    \label{fig:H2_form}
\end{figure}


\bsp	
\label{lastpage}
\end{document}